\documentclass{aa}  

\usepackage{graphicx}
\usepackage{txfonts}
\usepackage{placeins}
\usepackage{natbib,twoopt}
\usepackage[breaklinks=true]{hyperref} 
\bibpunct{(}{)}{;}{a}{}{,}             
\makeatletter
  \newcommandtwoopt{\citeads}[3][][]{\href{http://adsabs.harvard.edu/abs/#3}%
    {\def\hyper@linkstart##1##2{}%
     \let\hyper@linkend\@empty\citealp[#1][#2]{#3}}}
  \newcommandtwoopt{\citepads}[3][][]{\href{http://adsabs.harvard.edu/abs/#3}%
    {\def\hyper@linkstart##1##2{}%
     \let\hyper@linkend\@empty\citep[#1][#2]{#3}}}
  \newcommandtwoopt{\citetads}[3][][]{\href{http://adsabs.harvard.edu/abs/#3}%
    {\def\hyper@linkstart##1##2{}%
     \let\hyper@linkend\@empty\citet[#1][#2]{#3}}}
  \newcommandtwoopt{\citeyearads}[3][][]%
    {\href{http://adsabs.harvard.edu/abs/#3}
    {\def\hyper@linkstart##1##2{}%
     \let\hyper@linkend\@empty\citeyear[#1][#2]{#3}}}
\makeatother

\newcommand{\hi}{H\textsc{i}}
\newcommand{\hii}{H\textsc{ii}}
\newcommand{\oiii}[1]{O[\textsc{iii}]#1}
\newcommand{\spexxy}{\texttt{spexxy}}
\newcommand{\ULySS}{\texttt{ULySS}}
\newcommand{\ha}{\ensuremath{\mathrm{H}\alpha}}

\begin{document}

\title{The MUSE-Faint survey}

\subtitle{IV. Dissecting Leo T, a gas-rich relic with recent star formation}

\author{Daniel Vaz \inst{1}\fnmsep\inst{2}
  \and Jarle Brinchmann  \inst{1}\fnmsep\inst{2}\fnmsep\inst{3}
  \and Sebastiaan L. Zoutendijk  \inst{3}
  \and Leindert A. Boogaard \inst{4}
  \and Sebastian Kamann  \inst{5}
  \and Justin I. Read \inst{6}
  \and Martin M. Roth \inst{7}\fnmsep\inst{8}
  \and Peter M. Weilbacher  \inst{7}
  \and Matthias Steinmetz  \inst{7}
}
\institute{Instituto de Astrofísica e Ciências do Espaço, Universidade do Porto, CAUP, Rua das Estrelas, 4150-762 Porto, Portugal \\ \email{Daniel.Vaz@astro.up.pt}
  \and 
  Departamento de Física e Astronomia, Faculdade de Ciências, Universidade do Porto, Rua do Campo Alegre 687, PT4169-007 Porto, Portugal
  \and
  Leiden Observatory, Leiden University, PO Box 9513, 2300 RA Leiden, The Netherlands
  \and 
  Max Planck Institute for Astronomy, Königstuhl 17, 69117 Heidelberg, Germany
  \and 
  Astrophysics Research Institute, Liverpool John Moores University, IC2 Liverpool Science Park, 146 Brownlow Hill, Liverpool L3 5RF, United Kingdom
  \and 
  University of Surrey, Physics Department, Guildford, GU2 7XH, United Kingdom
  \and
  Leibniz-Institut für Astrophysik Potsdam (AIP), An der Sternwarte 16, 14482 Potsdam, Germany
  \and 
  Institut für Physik und Astronomie, Universität Potsdam, Karl-Liebknecht-Str. 24/25, 14476 Potsdam, Germany
  \\
}

\date{Received ; accepted }
   
 
\abstract
{Leo T ($M_V = -8.0$) is a peculiar dwarf galaxy that stands out for being both the faintest and the least massive galaxy known to contain neutral gas and to display signs of recent star formation. It is also extremely dark-matter dominated. 
As a result, Leo T presents an invaluable opportunity to study the processes of gas and star formation at the limit where galaxies are found to have rejuvenating episodes of star formation.}
{Our approach to studying Leo T involves analysing photometry and stellar spectra to identify member stars and gather information about their properties, such as line-of-sight velocities, stellar metallicities, and ages. By examining these characteristics, we aim to better understand the overall dynamics and stellar content of the galaxy and to compare the properties of its young and old stars.}
{Our study of Leo T relies on data from the Multi Unit Spectroscopic Explorer (MUSE) on the Very Large Telescope, which we use to identify 58 member stars of the galaxy. In addition, we supplement this information with spectroscopic data from the literature to bring the total number of member stars analysed to 75. To further our analysis, we complement these data with Hubble Space Telescope (HST) photometry. With these combined datasets, we delve deeper into the galaxy's stellar content and uncover new insights into its properties.}
{Our analysis reveals two distinct populations of stars in Leo T. The first population, with an age of $\lesssim 500~\mathrm{Myr}$, includes three emission-line Be stars comprising 15\% of the total number of young stars. 
The second population of stars is much older, with ages ranging from $>5~\mathrm{Gyr}$ to as high as $10~\mathrm{Gyr}$.
We combine MUSE data with literature data to obtain an overall velocity dispersion of $\sigma_{v} = 7.07^{+1.29}_{-1.12}~\mathrm{km\ s^{-1}}$ for Leo T. When we divide the sample of stars into young and old populations, we find that they have distinct kinematics. 
 Specifically, the young population has a velocity dispersion of $2.31^{+2.68}_{-1.65}\,\mathrm{km\ s^{-1}}$, contrasting with that of the old population, of $8.14^{+1.66}_{-1.38}\,\mathrm{km\ s^{-1}}$. 
 The fact that the kinematics of the cold neutral gas is in good agreement with the kinematics of the young population suggests that the recent star formation in Leo T is linked with the cold neutral gas. 
 We assess the existence of extended emission-line regions and find none to a surface brightness limit of~$< 1\times 10^{-20}\,\mathrm{erg}\,\mathrm{s}^{-1}\,\mathrm{cm}^{-2}~\mathrm{arcsec}^{-2}$ which corresponds to an upper limit on star formation of $\sim 10^{-11}~\mathrm{M_\odot~yr^{-1}~pc^{-2}}$, implying that the star formation in Leo T has ended.}
{}

\keywords{Spectroscopy, Galaxies, Leo T, Stars, Kinematics, Star Formation, Be Stars}

\maketitle
%
\section{Introduction}

Recent deep-imaging surveys have proven to be highly effective in
identifying faint companions to the Milky Way and other nearby galaxies
\citep{Simon_2019}. These faint galaxies have helped to elucidate a
number of challenges for the prevailing cosmological paradigm of
structure formation ---the $\Lambda$CDM--- on small scales; for example, collapsed
objects with $ M \leq 10^{11}\ \mathrm{M_\odot}$ \citep[e.g.][]{Bullock_2017}.

These challenges are closely related to the nature of dark matter.
Of particular interest here is the generic prediction in
$\Lambda$CDM that, in the absence of baryonic effects, dark matter halos
are expected to have cuspy density profiles at small scales, $\rho(r)
\propto 1/r$. However, observations in classical dwarfs often
indicate a cored density profile \citep{Walker_2011}.

The cores in these dwarfs might have been formed through baryonic
processes (e.g. supernova energy redistributes dark matter and creates cores \citealt{Navarro_1996, Bullock_2017}), but, as one progresses to ever fainter galaxies, the baryonic processes may act to terminate all star formation in the dark
matter halos due to the low binding energy compared to supernova
ejecta energy. The ultimate consequence is that the dark matter
profiles in these systems are expected to be more closely related to a
dark-matter-only profile
\citep[e.g.][]{Walker_2011,Di_Cintio_2013,Brook_2015,readDarkMatterCores2016a,Read_2019, Orkney_2021}. 
While a combination of stellar feedback and late minor mergers can still lower the inner dark matter density, 
the dark matter density profile remains cuspy even in the most extreme cases, 
with a central density that is lower by just a factor of approximately two \cite{Orkney_2021}.
At the same time, the shallow potential wells and potentially simple star formation histories also make these very faint galaxies
interesting laboratories to study galaxy formation at the smallest scales \citep{Jeon_2017, Bullock_2017, Rey_2020, Applebaum_2021, Gutcke_2022}.

The sample of ultrafaint dwarfs (UFDs) has grown rapidly in
recent years \citep[see][for a review]{Simon_2019} and we now have
a substantial sample with stellar velocity dispersions and
stellar population constraints. \citet{Simon_2019} argued that
a reasonable definition of UFDs is that they have
$M_V<-7.7$ ($L~\approx~10^5~\mathrm{L}_{\odot}$) and we adopt
this definition here. 
As such, our object of study is in the
transition from traditional dwarf spheroidal galaxies (dSph) to the
UFD class. 

Among the members of the faint and ultrafaint dwarf samples, Leo T has received more attention than any other.
The reason is simple: it is the faintest and least massive dwarf known to contain neutral gas and to show
signs of relatively recent star formation, which distinguishes it from the general UFD population. 
It is a natural `rosetta stone' for testing galaxy formation models, which have struggled to reproduce Leo T-like galaxies until very recently \citep{Rey_2020, Applebaum_2021, Gutcke_2022}. 
More detailed observations of Leo T will ultimately help us to further test the new models and to determine whether or not we are on the path towards a comprehensive and predictive theory of galaxy formation.

Leo T was discovered using SDSS imaging by \cite{Irwin_2007} and used
to be considered one of the brightest UFDs. With $M_V=-8.0$ \citep{Simon_2019}, it is
now just above the adopted brightness limit and therefore we choose to view it as a
transition object. 

The star formation history (SFH) of Leo T has been extensively studied
\citep{Irwin_2007,de_Jong_2008,Weisz_2012,Clementini_2012}. The studies broadly
agree that 50\% of the total stellar mass was
formed prior to $7.6~\mathrm{Gyr}$ ago, with the star formation beginning over 10 Gyr ago and continuing until recent times.
These latter authors found evidence of a quenching of star formation in Leo T about $25~\mathrm{Myr}$ ago. 
Interestingly, despite the extensive SFH, none of these studies found evidence
of an evolution in isochronal metallicity, such that over the
course of its lifetime it is consistent with a constant value of
$[M/H]\sim~-1.6$.

As stated above, Leo T is also known to contain neutral gas.
\cite{RyanWeber2008} and \cite{Adams2018}
concluded that Leo T contains both a cold neutral medium (CNM; $T\sim~800$K) and a warm neutral medium (WNM; $T\sim~6000$K) \hi. 
Specifically, the CNM was found to have kinematics different from those of the WNM (see Table~\ref{tab:LeoTP}). 
In addition to having a lower velocity dispersion, it has a lower mean velocity by about $2~\mathrm{km\ s^{-1}}$ (again, see Table~\ref{tab:LeoTP}).
A point of interest is that \hi\ gas dominates the baryon budget, being approximately twice as massive as the stellar component.

The only previous spectroscopic observations of Leo T are those of \cite{Simon_2007}, who used Keck/DEIMOS spectroscopy to
identify 19 stars as members of Leo T. These authors found a radial velocity of $v_{rad} = 38.1 \pm 2\ \mathrm{km\ s^{-1}}$ and a velocity dispersion of
$\sigma_{v_{rad}} = 7.5 \pm 1.6\ \mathrm{km\ s^{-1}}$. The stellar spectra indicated a metallicity of $\mathrm{[Fe/H]}=-2.29 \pm 0.15$, which was later revised in \cite{Kirby_2008, Kirby_2011, Kirby_2013}, with the latter indicating $\mathrm{[Fe/H]}=~-1.74~\pm~0.04$, by considering a weighted mean of the member stars. More recently, \cite{Simon_2019} reanalysed the same measurements but modelled the distribution as a Gaussian, fitting the most likely value of the mean. Their study suggests a metallicity value of $\mathrm{[Fe/H]}=~-1.91^{+0.12}_{-0.14}$.

From stellar kinematics, the estimated total
mass is of $8.2 \pm 3.6 \times 10^6 \mathrm{M_\odot}$, corresponding to a
mass-to-light ratio of $138 \pm 71\ [\mathrm{\frac{M_\odot}{L_\odot}}]$.  
\cite{Wolf_2010} included the spectroscopic data of Leo T in their derivation of an
accurate mass estimator for dispersion-supported stellar
systems. Combining with photometric results of
\cite{de_Jong_2008}, they derived a dynamical mass of
$M_{1/2} = 7.37^{+4.84}_{-2.96} \times 10^6\mathrm{M_\odot}$ and a
mass-to-light ratio of $\Upsilon^{V}_{1/2} = 110^{+70}_{-40}\
\mathrm{[\frac{M_\odot}{L_{V,\odot}}]}$ for Leo T.

The existing results show Leo T to be a transition object between
classical dwarfs and UFDs. However, the sample of stars
published before the observations reported here was too meagre to allow
strong constraints to be placed on the dark matter density profile or
to explore dynamical differences between different stellar
populations. 

In this paper, we present spectroscopic observations of Leo T using the Multi-Unit Spectroscopic Explorer
\citep[MUSE][]{baconMUSESecondgenerationVLT2010} integral field spectrograph (IFS). This is part of the MUSE-Faint survey of UFDs
\citep{Zoutendijk_2020,2021A&A...651A..80Z,2021arXiv211209374Z}. The data presented in the present paper were used by~\citet{2021arXiv211209374Z} to derive the density
profiles of Leo T  as part of a larger study of five faint and ultrafaint dwarfs, and also by~\citet{Regis_2021}
to place constraints on the abundance of axion-like particles.  In this paper, we also describe the derivation of the kinematic sample of Leo T member stars previously presented by~\citet{2021arXiv211209374Z}.

Here we present the data in more detail and use them to extend the analyses cited above.
We aim to densely map the stellar content, look for extended emission-line sources, and use the stellar spectra to measure the stellar metallicity and stellar kinematics. 
We search for identifiers of a young population, namely Be stars \citep[see][for a review]{porterClassicalBeStars2003, Rivinius_2013}: in low-metallicity environments, the evolution of young massive stars (mainly main sequence B stars, but also late O and early A stars) is thought to be affected by rotation \citep{Rivinius_2013, Schootemeijer_2022}. In cases of rapid rotation, the structure and appearance of the stars can be affected. Of relevance, they can acquire a decretion disk that will cause Balmer line emission around the star \citep{Rivinius_2013} that should be observable in our analysis.
The identification of Be stars in our sample would support the argument that these stars are common in low-metallicity environments and would confirm that Leo T has had recent star formation.
In such a case, we can split the stars by age using public HST/ACS data, and study the dynamical state of the young and old stars in the galaxy. 

This paper is organised as follows.
In Section~\ref{sec:observations} we present the data used, summarising the known properties of Leo T, and describing our observations 
and the data reduction process.
In Section~\ref{sec:data_analysis} we detail and justify the methods and steps that we apply to analyse the data. 
We report the results of the analyses in Section~\ref{sec:results} and discuss them in Section~\ref{sec:discussion}.
We present our conclusions in Section~\ref{sec:conclusions}.

\section{Observations and data reduction}
\label{sec:observations}

\subsection{Target: Leo T}
Table \ref{tab:LeoTP}  summarises the known properties of Leo T gathered from the literature. In the cases where multiple
studies on the same subject exist, we opted to select the results from the most recent study that takes into account previous results. 
In the case of metallicity, we adopted $[\mathrm{Fe/H]} \sim -1.6$ as all photometric studies are consistent with this value. We point out that this value is not in agreement with the metallicity indicated by \cite{Simon_2019} but is in reasonable agreement with the spectroscopic measurement of \cite{Kirby_2013}, which indicates $\mathrm{[Fe/H]}=-1.74~\pm~0.04$

Relevant here is the fact that \cite{Adams2018} found a difference in kinematics between the components of warm and cold
neutral gas, which we address in our analysis below.

\begin{table}[]
\centering
\caption{Selected Leo T properties gathered from the existing literature. }
\begin{tabular}{lccrr}
\hline\hline
\multicolumn{1}{c}{Property}                                     & \multicolumn{1}{c}{Value}                    \\ \hline
$\alpha_{J2000}$                                     & $09\ 34\ 53.4$             \\
$\delta_{J2000}$                                 & $+17\ 03\ 05$             \\
Distance $\mathrm{[kpc]^{(1)}}$                                 & $409^{+29}_{-27}$             \\
Luminosity $\mathrm{[L_{\odot,V}]^{(2)}}$                       & $1.4 \times 10^5$                  \\
3D $r_{1/2}~\mathrm{[pc]^{(2)}}$                                         & $152 \pm 21$                       \\
2D $r_{1/2}~\mathrm{[pc]^{(2)}}$                                   & $115 \pm 17$                       \\
$v_\star$ $\mathrm{[km\ s^{-1}]^{(3)}}$                       & $38.1 \pm 2$                       \\
$\sigma_\star$ $\mathrm{[km\ s^{-1}]^{(3)}}$                  & $7.5 \pm 1.6$                      \\
$\mathrm{M_{1/2}}$ $\mathrm{[M_\odot]^{(2)}}$                                & $7.37^{+4.84}_{-2.96} \times 10^6$ \\
$\Upsilon^{V}_{1/2}$ $\mathrm{\left[\frac{M_\odot}{L_{V,\odot}}\right]^{(2)}}$ & $110^{+70}_{-40}$                  \\
$\mathrm{M_{HI}}$ $\mathrm{[M_\odot]^{(4)}}$                                 & $4.1 \pm 0.4 \times 10^5$          \\
$v_{\mathrm{CNM}}$ $\mathrm{[km\ s^{-1}]^{(4)}}$                       & $37.4 \pm 0.1$                     \\
$v_{\mathrm{WNM}}$ $\mathrm{[km\ s^{-1}]^{(4)}}$                       & $39.6 \pm 0.1$                     \\
$\sigma_{\mathrm{CNM}}$ $\mathrm{[km\ s^{-1}]^{(4)}}$                  & $2.5 \pm 0.1$                      \\
$\sigma_{\mathrm{WNM}}$ $\mathrm{[km\ s^{-1}]^{(4)}}$                 & $7.1 \pm 0.4$                      \\
$\mathrm{[Fe/H]}^{(5)}$                                             & $\sim -1.6$                        \\
\hline
\end{tabular}
\tablebib{$^1$\citet{Clementini_2012}; $^2$\citet{Wolf_2010};
  $^3$\citet{Simon_2007}; $^4$\citet{Adams2018};
  $^5$\citet{Weisz_2012}.}
        \label{tab:LeoTP}
\end{table}

\subsection{Observations}
The central region of Leo T was observed as part of MUSE-Faint
\citep{Zoutendijk_2020}, a MUSE GTO survey of UFD
galaxies (PI Brinchmann). MUSE \citep{baconMUSESecondgenerationVLT2010} is a
large-field medium-spectral-resolution integrated field spectrograph installed at
Unit Telescope 4 of the Very Large Telescope (VLT). For the
observations of Leo T, we used the wide-field mode with
ground-layer adaptive optics (WFM-AO), which
provides a $1 \times 1 \; \mathrm{arcmin^2}$ field of view (FoV) split
into 24 slices, each sent to a separate integral field unit (IFU).
This configuration enables a $ 0.2\;\mathrm{arcsec\,pixel^{-1}}$
spatial sampling and a wavelength sampling of $1.25$ \AA\,
$\mathrm{pixel^{-1}}$, with a nominal wavelength coverage of
$4700-9350$ \AA\ with the range 5807--5963 \AA\ taken out by a notch
filter to avoid contamination by the sodium lasers used in AO mode.

\begin{table}[ht]
\footnotesize
\centering
\caption{List of the MUSE observations of Leo T. We provide information of the exposure time, grade, and FWHM at 7000 \AA.
Each observation corresponds to three exposures of 15 minutes. The grade indicates whether the observing conditions fit within the constraints. An observation made fully within the constraints has grade A and one made mostly within the constraint has grade B.}
\begin{tabular}{lccccrr}
    \hline\hline
    \multicolumn{1}{c}{Date} & \multicolumn{1}{c}{Exp. Time} & \multicolumn{1}{c}{Grade} & \multicolumn{1}{c}{FWHM @ 7000\AA} \\ 
                             & \multicolumn{1}{c}{[min.]}    &                           & \multicolumn{1}{c}{[arcsec]} \\
    \hline
    2018-02-14    & 45                  & A              & 0.52 \\
    2018-03-17    & 45                  & A              & 0.53 \\
    2018-04-16/17 & 45                  & A              & 0.69 \\
    2019-01-06/07 & 45                  & B              & 0.96  \\
    2019-04-09    & 45                  & A              & 0.60  \\
    \hline
    \end{tabular}
\label{tab:obs}
\end{table}

The final reduced cube is a combination of a total of 3.75 hours taken
under programme IDs 0100.D-0807, 0101.D-0300, 0102.D-0372, and
0103.D-0705 between 14 February 2018 and 9 April 2019. The 15
exposures, all of which had a duration of 900s, are detailed in
Table~\ref{tab:obs}, which provides with their full width at half maximum (FWHM).

\subsection{Data reduction}
The data were processed with the ESO Recipe Execution Tool (EsoRex;
version 3.13) using MUSE Data Reduction Software (DRS; version
2.8.1; \citealt{weilbacher2020data}). The reduction followed standard
procedures: first, for each science exposure, we applied a bad-pixel table
created by \cite{Bacon2017}. We then applied the relevant master
bias, master flat, trace table, wavelength calibration table, geometry
table, and twilight cube to subtract the bias
current, corrected for pixel-to-pixel sensitivity variations, and calibrated each pixel in position, wavelength, and
flux. We also applied an illumination exposure to account for flat
field variations due to temperature differences between the science
and calibration exposures.

The resulting 24 pixel tables were corrected for atmospheric refraction
and a flux calibration and telluric correction was applied. We
enabled autocalibration using the deep-field method, an updated
version of the method described by \cite{Bacon2017} and included in the DRS. We corrected for emission
lines caused by Raman scattering from the AO lasers and applied sky subtraction. Furthermore, we corrected the wavelengths
for barycentric motion.  The exposures were spatially aligned and the data were finally combined into one pixel table and one data cube.
To remove residual sky signatures present in the data cube, we used the Zurich Atmosphere Purge \citep[ZAP; version 2.1;][]{Soto_2016}.

After completing the data reduction, we inspected the final result and detected a bright streak across the exposure taken on 16 April 2018, which we identified as a satellite track. We corrected this by manually masking out the affected pixels.

\subsection{Extracting spectra}

\begin{figure*}[t]
\centering
\includegraphics[scale=0.7]{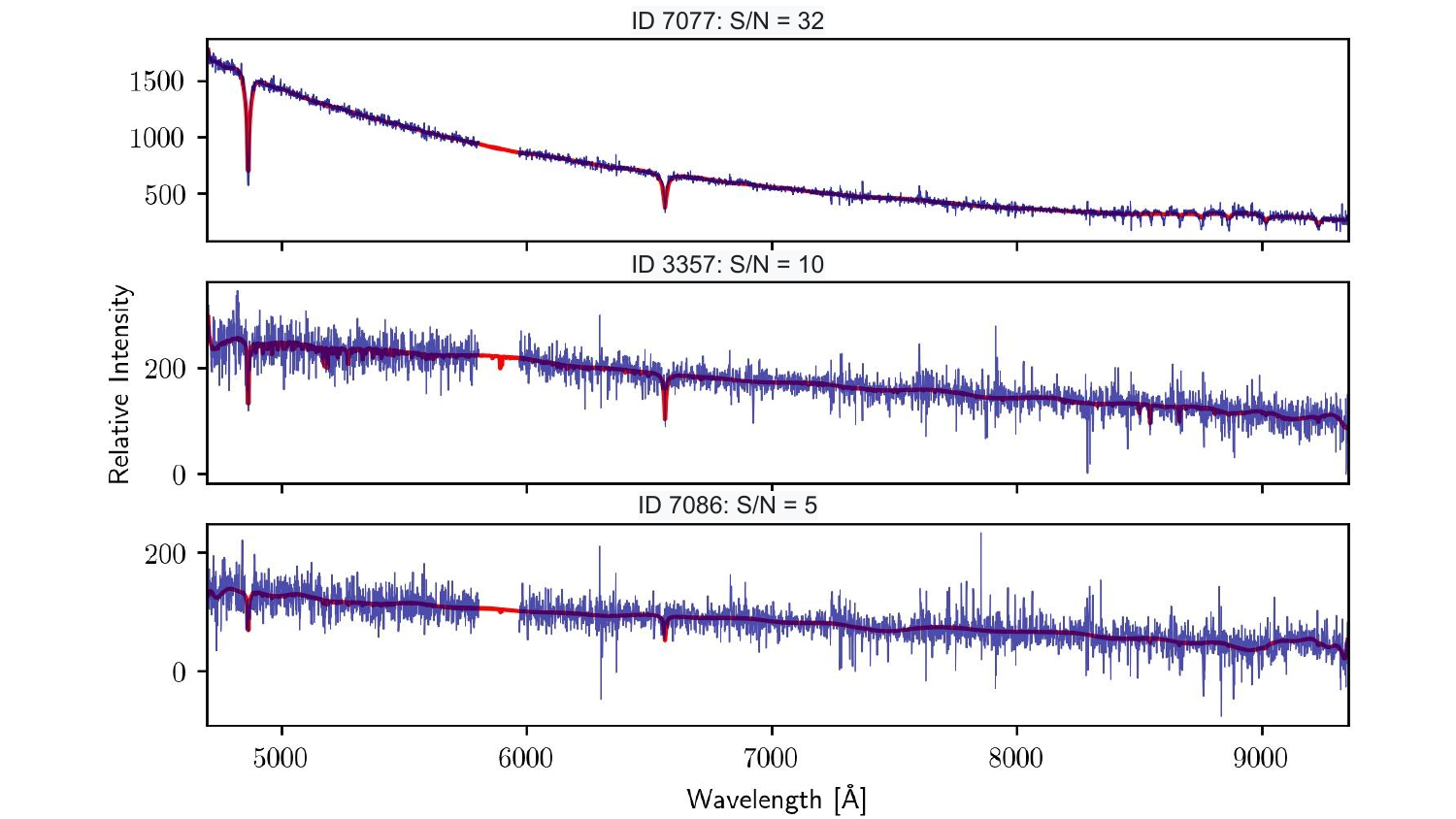}
\caption{Examples of spectra and their corresponding \spexxy\ best fit, using high-, medium-, and low-S/N spectra for illustrative purposes. The ID number is a running number provided by SExtractor.}
        \label{fig:spec_examples}
\end{figure*}

The last step in data processing is to extract spectra from the
sources present in the data cube.  We used PampleMuse
\citep{Kamann_2012}, which is optimised for the  extraction
 of stars from integral-field spectroscopic
observations of crowded stellar fields. This software requires a
high-fidelity input source catalogue (e.g. from HST data) to identify
and locate sources in the data cube. For this, we created a master
list of sources from public HST Advanced Camera for Surveys (ACS) data
from Leo T\footnote{HST Proposals 12914, Principal Investigator Tuan Do
and 14224, Principal Investigator Carme Gallart}. We used SExtractor \cite{Bertin1996SExtractorSF} to
detect sources in the field.

Succinctly, PampelMuse works as follows: it takes the source list to optimise positions and makes a
first estimate of the point-spread function (PSF) using a subset of
the sources present in the data cube. We adopted a Moffat function for
the PSF and measured an FWHM of $3.05\;\mathrm{pixels}$ or
$0.61\;\mathrm{arcsec}$ at a wavelength of $7000$ \AA\ in the final
combined cube. Armed with this PSF, an optimally extracted spectrum is
obtained for each sufficiently bright source. This
procedure led to the extraction of spectra for 271 sources. We
manually inspected the spectra and identified 19 as clearly
being spectra of background galaxies, which were excluded from the analyses. Of relevance, we also
identified three emission line stars, the spectra and scientific meaning of which we discuss below.

Figure~\ref{fig:spec_examples} shows three examples of the extracted spectra. These spectra correspond to stars that were identified as members of Leo  T. Therefore, we also plot the best fit performed by \spexxy\, as described in the following section. The spectra shown here were selected according to their signal-to-noise ratio (S/N): we show the spectrum with the highest S/N in the sample, one with intermediate S/N, and one with low S/N.

\section{Data analyses}
\label{sec:data_analysis}

The spectra extracted through the process discussed above consist of those for
both member and non-member stars of Leo T. Furthermore, not all
are suitable for analysis.
In this section, in addition to describing the methods that we use to analyse the data, we present the steps
that we took to determine which stars are likely to belong to Leo T and which can be used to obtain reliable measurements of physical parameters.

\subsection{Measuring stellar velocities and abundances}
\label{sec:spexxy}

The spectra that we extracted from the reduced data cube, as a
general rule, have a modest S/N and spectral resolution ($R\sim 3000$). 
As such, we adopt the same procedure as in
previous articles in the series \citep{Zoutendijk_2020}. 

We use the \spexxy\ full-spectrum fitting code \citep{gup-87}
for the estimation of the physical parameters. This allows us to
obtain line-of-sight velocity, metal abundance, effective temperature,
and surface gravity for all stars, as long as the S/N is sufficient.

\spexxy\ works by performing an interpolation over a grid
of PHOENIX \citep{Husser_2013} model spectra, which parameterizes
effective temperature, surface gravity, iron abundance ([Fe/H]), and
alpha-element abundance. Because the quality of the spectra is insufficient to allow distinction between different values of the alpha-element abundance ratio, we fixed the alpha-element abundance ratio to
solar and left the other parameters free to be fitted. 
We proceeded this way in order to directly compare our measurements of [Fe/H] with our estimates of metallicity ([M/H]) using isochrone fitting (see Section~\ref{sec:mem}). It is worth noting that the equality [M/H]~=~[Fe/H] only holds when the ratio of alpha elements to iron ---denoted [$\alpha$/Fe]--- is zero. In the event that the stars in Leo T are enriched in alpha elements, we would expect to observe [M/H] > [Fe/H].

With this procedure, we were able to successfully fit 130 spectra, all of which have apparent magnitude $< 24$. 
Within the sample that was not successfully fitted, we found three emission line stars. Because the S/N values for the three spectra are 
sufficiently high, we concluded that their spectral type is not covered by the PHOENIX grid we use, which would lead to an unsuccessful fit. 
In this case, and to obtain line-of-sight velocity measurements for these stars, we adopted \ULySS\ \citep{Cappellari04, Koleva_2009} as an alternative to obtain radial velocities. We describe \ULySS\ in Appendix \ref{sec:uly}, and compare with the results we obtain when using \spexxy. However, as the results are consistent with each other, we adopt \spexxy\ as our default method to determine the radial velocities, with the exception of the three emission line stars.

\subsection{Selecting members}\label{sec:mem}

When selecting the Leo T member stars in our sample, we must take into account a possible contamination by Milky Way stars.
As such, we estimate how many stars in our FoV are likely to be Milky Way members using the latest available Besançon model of the Milky Way \citep{2014A&A...569A..13R,2012A&A...538A.106R,2003A&A...409..523R}.
We simulated Milky Way star counts in the direction of Leo T and within
a solid angle of $1~\mathrm{deg^2}$. In the simulation, we get a total of
3083 stars when only considering apparent magnitudes between 20 and 24, which is the range of our sample.
This corresponds to $\approx 0.86$ stars in the MUSE FoV that would make it into our sample.
Therefore, and from the model, we find it reasonable to expect that approximately one of the Leo T member candidates is actually a member of the Milky Way.

\begin{figure}[]
\centering
\includegraphics[scale=0.6]{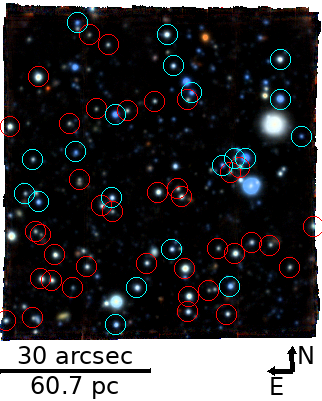}
\caption{Composite color image of Leo T based on data from the MUSE-Faint survey, created with fits2comp\protect\footnotemark. We used the Johnson-Cousins filters I, R, and V to create the red, green, and blue channels of the image, respectively. The 58 stars identified in this paper as Leo T members that are located within the bounds of MUSE-Faint observations are identified with circles. As we identify two stellar populations (see Section~\ref{sec:ages}), we use blue and red circles for the younger and older population, respectively. The angular and physical scales of the image are indicated in the lower left corner. Directions north and east are indicated in the lower right corner of the image.}
        \label{fig:whitelight}
\end{figure}

Furthermore, we took into account the S/N of the spectra when selecting our sample of Leo T member stars.
As \spexxy\ has been found to underestimate velocity uncertainties for spectra with a low S/N
\citep{kamannMUSECrowdedField2016}, we only considered spectra with a median S/N$>3$, per $1.5\AA$, similar to the approach taken in \cite{Zoutendijk_2020}. Of the 130 spectra successfully fitted with \spexxy, 56 satisfy this criterion. 
In addition, as the three emission-line stars also satisfy this criterion, they were also included in the sample.

Finally, in order to detect possible contamination, namely from Milky Way stars, as stated above, we performed a photometric analysis. To this end, we used F606W and F814W photometry of the public HST/ACS data, the same used to create a master list of sources.
Using the photometry data, we drew a colour--magnitude diagram and matched it to isochrones from PARSEC stellar tracks and
isochrones \citep{Bressan_2012}. 
After trying different values, we assume a fixed metallicity of
$\mathrm{[M/H}]=-1.6$, which is consistent with the value of \cite{Weisz_2012} and is also in reasonable agreement with the value found by \cite{Kirby_2013}. 
We find that this value for [M/H] leads to the best fit for our MUSE data, and we discuss this further in Section~\ref{sec:ages}.

We adopted $A_V = 0.1$ magnitudes for the interstellar extinction as a way to better align the isochrones with the data. This value is in good agreement with \cite{SFD1998}, from which we get $A_{V_{SFD}} = 0.0959$.
We find that only one of the 59 stars is inconsistent with the isochrones (having $F606W-F814W\approx2$) and was excluded. 
The estimated line-of-sight velocity for this star is $\mathrm{v_{los}} = -31.7 \pm 5.5\ \mathrm{km\ s^{-1}}$. 
We consider this most likely to be a foreground M-dwarf Milky Way interloper star, which is perfectly in line with the Besançon model prediction.

Thus, we obtained a sample of 58 stars that we consider to be plausible members of Leo T. 
In Figure~\ref{fig:whitelight} the positions of these stars are marked on the MUSE FoV. The colour--magnitude diagram referred to above, which includes these 58 stars, is shown in Figure~\ref{fig:color}. 
Upon  further exploration of this diagram, we identified two different populations: a younger population, with ages below $1\ \mathrm{Gyr}$, and an 
older population with ages above $5\ \mathrm{Gyr}$, which is consistent with ages as high as $10\ \mathrm{Gyr}$. 

To make this assignment, we compared the colour--magnitude diagram of stars with a grid of isochrones based on previous results in the literature. 
We used PARSEC stellar tracks and isochrones \citep{Bressan_2012} to generate 10 isochrones spaced equally between 0.1 and 1 Gyr, and 7
isochrones spaced equally between 5 and 11 Gyr, all with fixed metallicity of $\mathrm{[M/H}]=-1.6$. 
We assign each star the age of the nearest isochrone in the colour--magnitude diagram. Within our sample, we identify stars whose best fit is for ages  as low as $200$~Myr and others for ages as high as $10$~Gyr. 
Due to the high age diversity in our sample, we opted to divide the stars into two samples: stars whose best fit is for ages $\leq 1$ Gyr and stars whose best fit is for ages $\geq 5$ Gyr.

When we fix the metallicity, there is a degeneracy between the different isochrones in certain parts of the colour--colour space. This translates into an uncertain age assignment for stars falling in this region (darker-blue colored in Figure~\ref{fig:color}).
In order to assess the sensitivity of our results below to the classification of stars in this region, we repeated the analysis while changing the assignment of stars to the young or old isochrones in this region. This did not significantly change our results, and therefore the discussion below is robust to the classification of stars in this degenerate region.

\footnotetext{\url{https://github.com/slzoutendijk/fits2comp}} 

\subsection{Distributions}
\label{sec:dist}

While the methods described above are designed to analyse each star individually, here we describe the methods that we use to characterise Leo T as a whole.
To this end, we adopt a Markov Chain Monte Carlo (MCMC) method following the approach of~\citet{Zoutendijk_2020} in order to estimate the intrinsic mean 
and standard deviation of the velocity and metallicity distributions of Leo T. 
Here we use the logarithmic abundance relative to solar, [Fe/H], as our metallicity variable.

Here, the main assumption we make is that the velocity and metallicity distributions of the Leo T members can be
well described by a Gaussian distribution. Also, we consider the possibility that there might be a contaminating
population of stars belonging to the Milky Way that were not detected and whose distribution we take to be uniform in velocity across the velocity range considered, similarly to~\protect\citet{Martin_2018}. 

We describe the MCMC method as follows. If we denote the likelihood that star $i$  belongs to Leo T as $m_i$, the global
likelihood is given by
\begin{equation}
\begin{aligned}
  \label{eq:likelihood}
  \mathcal{L}(\mu_{int}, \sigma_{int} | x_i, \epsilon_i) = 
  \displaystyle\prod_{i} \left[ m_i \frac{ 1 }{\sqrt{2\pi}
      \sigma_{obs,i}} \exp \left(-\frac{1}{2} \left(\frac{x_i -
          \mu_{int}}{\sigma_{obs,i}} \right)^2\right) \right. \\ \left. + \frac{1-m_i}{\mu_{bg}^{MAX} - \mu_{bg}^{MIN}} \right],
  \end{aligned}
\end{equation}
where $\mu_{int}$ and $\sigma_{int}$ are the intrinsic mean value and intrinsic dispersion of both Leo T parameters being studied. 
The value $\sigma_{obs,i} = (\sigma_{int}^2 + \epsilon_i ^2)^{1/2}$ is the observed velocity--metallicity dispersion for each star $i$, while
$x_i$ denotes the measurement and $\epsilon_i$ the measurement uncertainty of the parameter.  

For the background population, we fix the distribution by setting an upper and lower limit on the quantity considered. For the velocity, we consider $\mu^{\mathrm{MIN, MAX}}_{\mathrm{bg}} = 20, 70\,\mathrm{km}\,\mathrm{s}^{-1}$ and for the metallicity $\mu^{\mathrm{MIN, MAX}}_{\mathrm{bg}} = -2.5, 0$. 
For $\mu_{\mathrm{int}}$, we take a flat prior between $20$ and $70$$\,\mathrm{km}\,\mathrm{s}^{-1}$ for velocity and between $-2.5$ and $0$ for [Fe/H]. We also use a flat prior on $\sigma_{\mathrm{int}}$ between $0$ and $30\,\mathrm{km}\mathrm{s}^{-1}$ and between $0$ and $1$ for velocity and [Fe/H], respectively. 
The membership likelihood is allowed to take values over the full range of 0 to 1 with a flat prior over that range, in both cases.
We use \texttt{emcee} \citep{FM2013} to infer the posterior constraints on the parameters.

\section{Results}
\label{sec:results}

\subsection{Stellar ages}\label{sec:ages}

\begin{figure*}[]
\resizebox{\hsize}{!} {\includegraphics[scale=0.6]{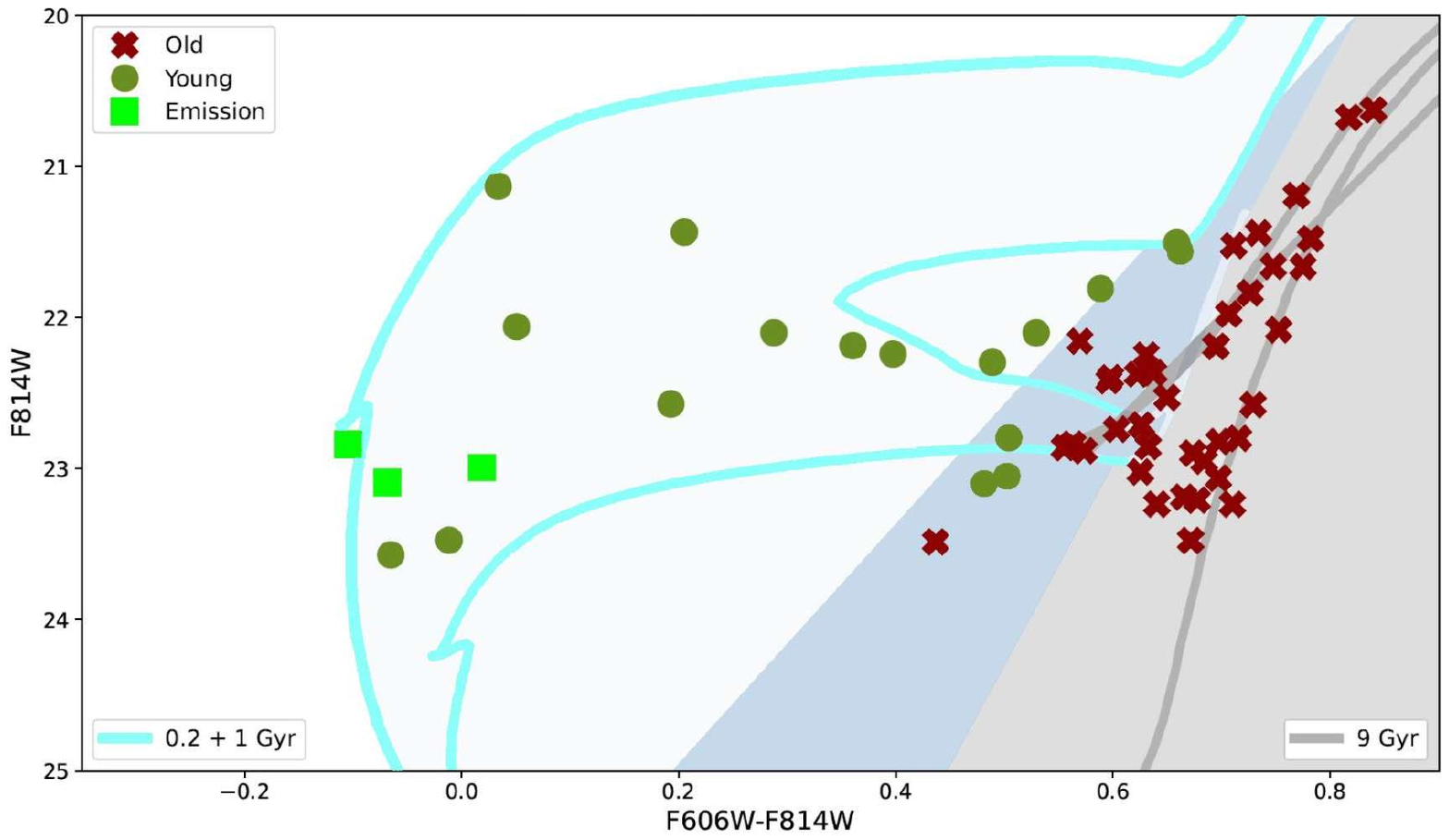}}
\caption{Colour--magnitude diagram of the 58 Leo T stars detected with
  MUSE, plotted against PARSEC isochrones drawn for constant $[\mathrm{Fe/H}] = -1.6$ and variable age. Magnitudes given are on the Vega Magnitude System.
  The region spanned by $0.1-1$ Gyr isochrones is shaded in light blue with two representative isochrones shown for 0.2 and 1.0 Gyr. 
  The region spanned by $>5\,\mathrm{Gyr}$ isochrones is shown in dark grey shading and a 9 Gyr isochrone is shown for illustration. 
  At the meeting point of the two regions there is an overlap represented by a region shown in darker blue; here there is considerable degeneracy between the isochrones.
  The stars that were found to be consistent with the younger isochrones are shown as dark green discs, with the emission line stars shown as green squares. 
  The stars consistent with the older isochrones are shown as red crosses.}
        \label{fig:color}
\end{figure*}

As mentioned in Section~\ref{sec:mem}, we used the colour--magnitude diagram of the 58 stars considered to be members of Leo T to probe for different stellar populations.
Figure \ref{fig:color} shows the colour--magnitude diagram for the 58 stars plotted against the PARSEC isochrones. The stars are marked with different
symbols of different colours corresponding to the age group to which the stars were assigned: red crosses correspond to older stars with ages $\geq 5$~ Gyr, while green circles correspond to younger stars with ages $\leq 1$ Gyr. The three emission-line stars are marked as green squares.

Although we divided the stars into two populations, the sample covers a wide range of ages, with some stars consistent with very old ages $> 10$~Gyr, while others appear to be very young, with ages of about $200$~Myr or younger. The three emission line stars are just a few examples of these very young stars, which we address in more detail below.  Moreover, a total of 20 stars are consistent with ages of $< 1$~Gyr. This is consistent with previous findings that Leo T had star formation in its recent history, making Leo T the faintest galaxy with evidence of recent star formation, partly justifying the
choice of \citet{Simon_2019} to set the cut-off point for UFDs just below the luminosity of Leo T.

Furthermore, all stars are consistent with a metallicity of $\mathrm{[M/H]} = -1.6$, which is in line with the existing photometric studies of Leo T
\citep[e.g.][]{de_Jong_2008,Weisz_2012,Clementini_2012}. 
We tried different values, but this is the one that maximises the number of stars that are consistent with the isochrones. 
Using this value, the older stars in the sample are consistent with ages of~$\sim~10~\mathrm{Gyr}$. This is in line with the findings of \cite{Clementini_2012}, which indicates this value as an upper limit for the stellar age on Leo T. Additionally, considering a lower value for metallicity (e.g. $\mathrm{[M/H]} = -1.9$) would make the redder stars in our colour--magnitude diagram inconsistent with the isochrones. 

On the other hand,  the value of $\mathrm{[M/H]} = -1.6$ also makes the emission-line stars consistent with being on the main sequence with ages of $<500$ Myr, which is consistent with the expectation that these stars have ages of $< 1$ Gyr. Once again, if we considered a lower value for metallicity, the best-fit isochrone for the emission line stars would correspond to an older age, which would start to be inconsistent with the expectations for these stars.

\subsection{Emission line stars}\label{sec:emstars}
\begin{center}
\begin{figure*}[]
\resizebox{\hsize}{!}{\includegraphics[scale=.75]{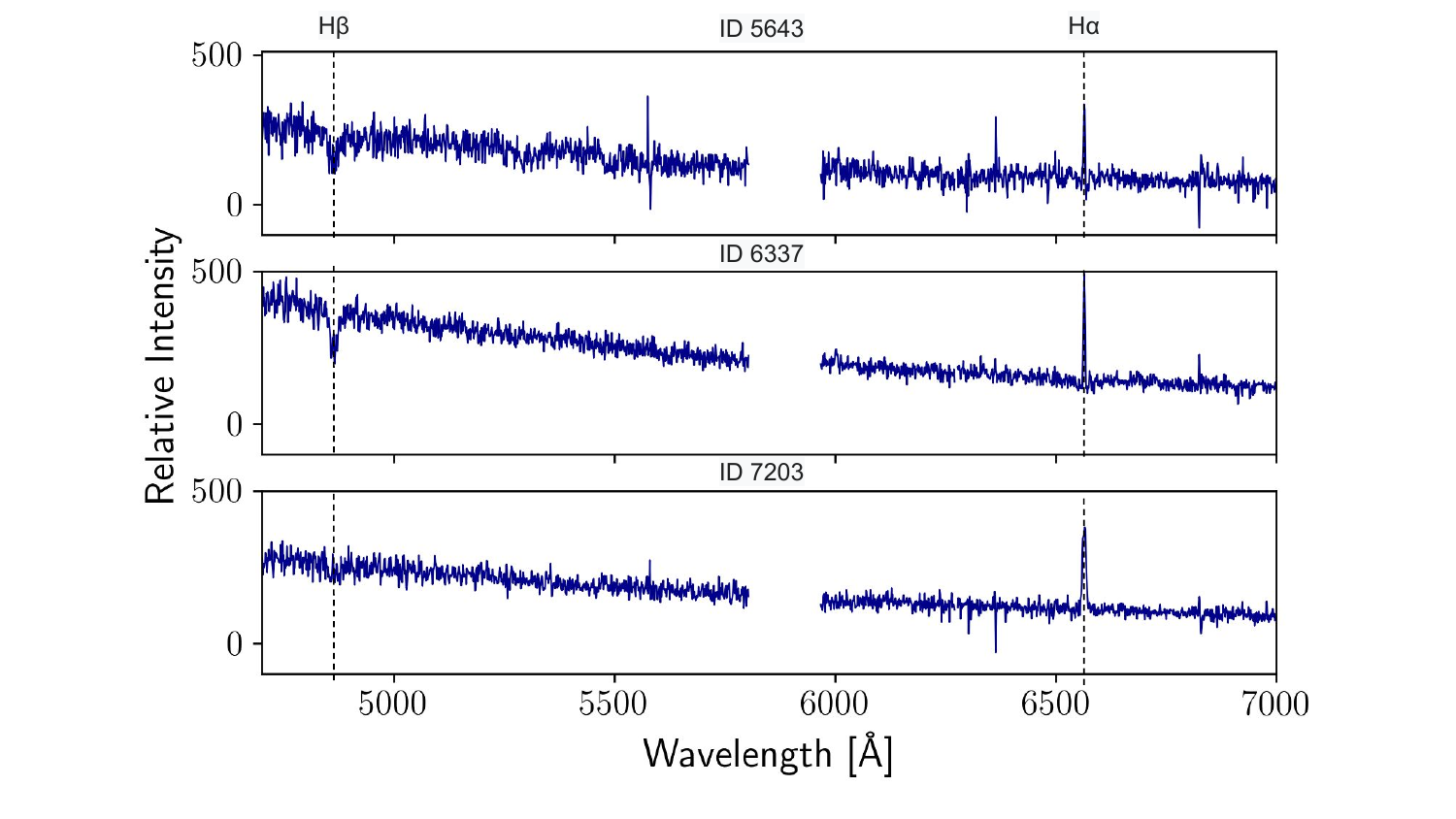}}
        \caption{Spectra of the three emission-line stars with the positions of the H$\alpha$ and H$\beta$ lines indicated. The ID number is a running number provided by
    SExtractor.} 
        \label{fig:emstars}
\end{figure*}
\end{center}

The stars marked with green squares in Figure \ref{fig:color} stand out not only for their youth but also for their spectra, which are
shown in Figure~\ref{fig:emstars}.  
All of them show clear H$\alpha$ emission on top of a blue stellar spectrum showing H$\beta$ in absorption.

 As mentioned above, our attempt to fit the spectra with \spexxy\ failed due to a lack of suitable templates. Therefore, we followed the approach of \cite{Roth_2018} using \ULySS\ \citep{Cappellari04, Koleva_2009} and the empirical \textit{MIUSCAT} library \citep{Vazdekis_2012} (see also Appendix~\ref{sec:uly}) to fit the stellar spectra of the three stars. We obtained the best fit for Be stars with an effective temperature of the order of
$T_{\mathrm{eff}} \sim 10^4$ K. 
It was also possible to fit their line-of-sight velocities, and these were used in the Leo T dynamics analyses discussed in Section \ref{sec:kin}.

On the basis of these results, we tentatively identify the stars as Be stars. Assuming that they are classic Be stars \mbox{\citep[e.g.][]{porterClassicalBeStars2003, Rivinius_2013}}, it is notable that they make up 15\% of the young stars in our sample and we return to
this subject in Section~\ref{sec:discussion} \footnote{We note that the literature tends to include early A emission stars under the classical Be star umbrella and we do not try to distinguish between early Ae and Be stars here.}. 

\begin{table*}
  \centering
    \caption{Properties of the emission-line stars in Leo T. The ID number is a running number provided by
    SExtractor. Positions and apparent and absolute magnitudes (Vega system) are taken from the HST imaging, and the \ha\ flux,
    equivalent width, and line width are from the MUSE data and are corrected for instrumental dispersion.}
  \begin{tabular}{lcccccccrr}
  \hline\hline
    \multicolumn{1}{c}{ID}
    & \multicolumn{1}{c}{$\alpha_{\mathrm{J2000}}$}
    & \multicolumn{1}{c}{$\delta_{\mathrm{J2000}}$}
    & \multicolumn{1}{c}{$m_{\mathrm{F606W}}$}
    & \multicolumn{1}{c}{$M_{\mathrm{F606W}}$}
    & \multicolumn{1}{c}{$m_{\mathrm{F814W}}$}
    & \multicolumn{1}{c}{$M_{\mathrm{F814W}}$}
    & \multicolumn{1}{c}{$F_{\ha}$}
    & \multicolumn{1}{c}{EW$_{\ha}$}
    & \multicolumn{1}{c}{$\sigma_{\ha}$}  \\
    & \multicolumn{1}{c}{[deg]} 
    & \multicolumn{1}{c}{[deg]} 
    & & & &
    & \multicolumn{1}{c}{[$10^{-20}
      \,\mathrm{erg}\,\mathrm{s}^{-1}\,\mathrm{cm}^{-2}$]}
    & \multicolumn{1}{c}{[\AA]} 
    & \multicolumn{1}{c}{[$\mathrm{km~s^{-1}}$]}  \\ \hline
    5643  &  143.7278113  & 17.0597494 & 23.02 &   $-0.04$ & 23.09 & $0.03$
    & $1089 \pm50$ & $-9.82\pm 0.85$  & $55.5\pm 4.6$ \\
    6337  &  143.7211871  & 17.0558321 & 22.73 & $-0.33$ & 22.84 & $- 0.22$
    & $1684\pm 34$ & $-12.12 \pm 0.38$ & $65.0 \pm 2.2$ \\
    7203  &  143.7194132  & 17.0517905 & 23.01 &   $- 0.05$  & 22.99 & $- 0.07$
    & $2925\pm 47$ & $-27.01\pm 0.47$ & $166.8\pm 3.4$ \\
    \hline
  \end{tabular}
  \label{tab:emlinestars}
\end{table*}

Table~\ref{tab:emlinestars} provides positions, absolute and apparent
magnitudes, and \ha\ line flux, equivalent width, and line
width. The existing near-infrared (NIR) data for Leo T that we are aware of are
not deep enough to detect the emission-line stars and therefore we are
unable to measure NIR excesses, which are commonly seen in these
stars \citep{porterClassicalBeStars2003}. We
note that the absolute magnitudes are towards the lower end of those
found by~\citet{Mathew_2008}, which is consistent with the star
population in Leo T being somewhat older than the majority surveyed in
that paper.

We also verified that the Balmer lines are unlikely to originate in a
faint ionised gas region around the Be stars. To this end, we ran some simple
Cloudy models \citep{1998PASP..110..761F} using B-star model atmospheres
from~\citet{Lanz_2007} with temperatures and
luminosities spanning from B9 to B0 stars. Calculations were performed
with version 17.01 of Cloudy, last described by
\cite{2017RMxAA..53..385F}. We find that, for stars with
$T_{\mathrm{eff}} < 24$ kK, the measured line fluxes are too large to
be explained by nebular emission from an ionised region around the
stars. Taken together with the fact that the best fit for the effective temperature for these spectra is $T_{\mathrm{eff}}<20$ kK, we conclude that a nebular origin for the lines is
unlikely. This is further substantiated by the finding that star ID 7203 has a substantial line width of 167 km/s, which is inconsistent with a nebular origin and is therefore most likely indicative of emission originating near a rapidly rotating star \cite[see][]{Schootemeijer_2022}

\subsection{Stellar metallicity}\label{sec:met}

\begin{figure}[]
\centering
\includegraphics[scale=.55]{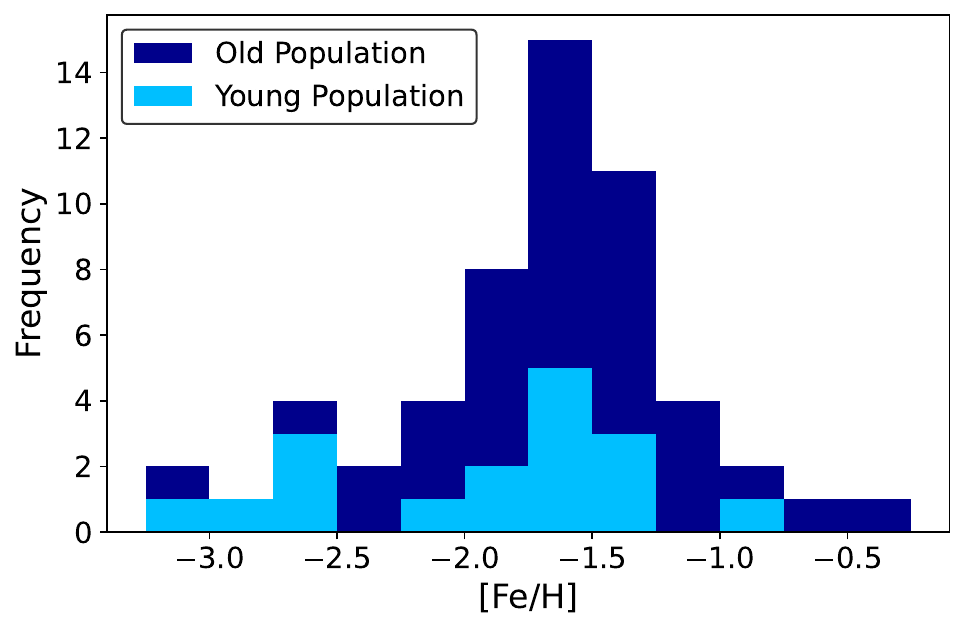}
\caption{Distribution of the metallicity ([Fe/H]) of 55 Leo T stars estimated using \spexxy. The younger population, consisting of 17 stars, is represented with a lighter colour.}
        \label{fig:histfe}
\end{figure}

\begin{figure}[]
\centering
\includegraphics[scale=0.45]{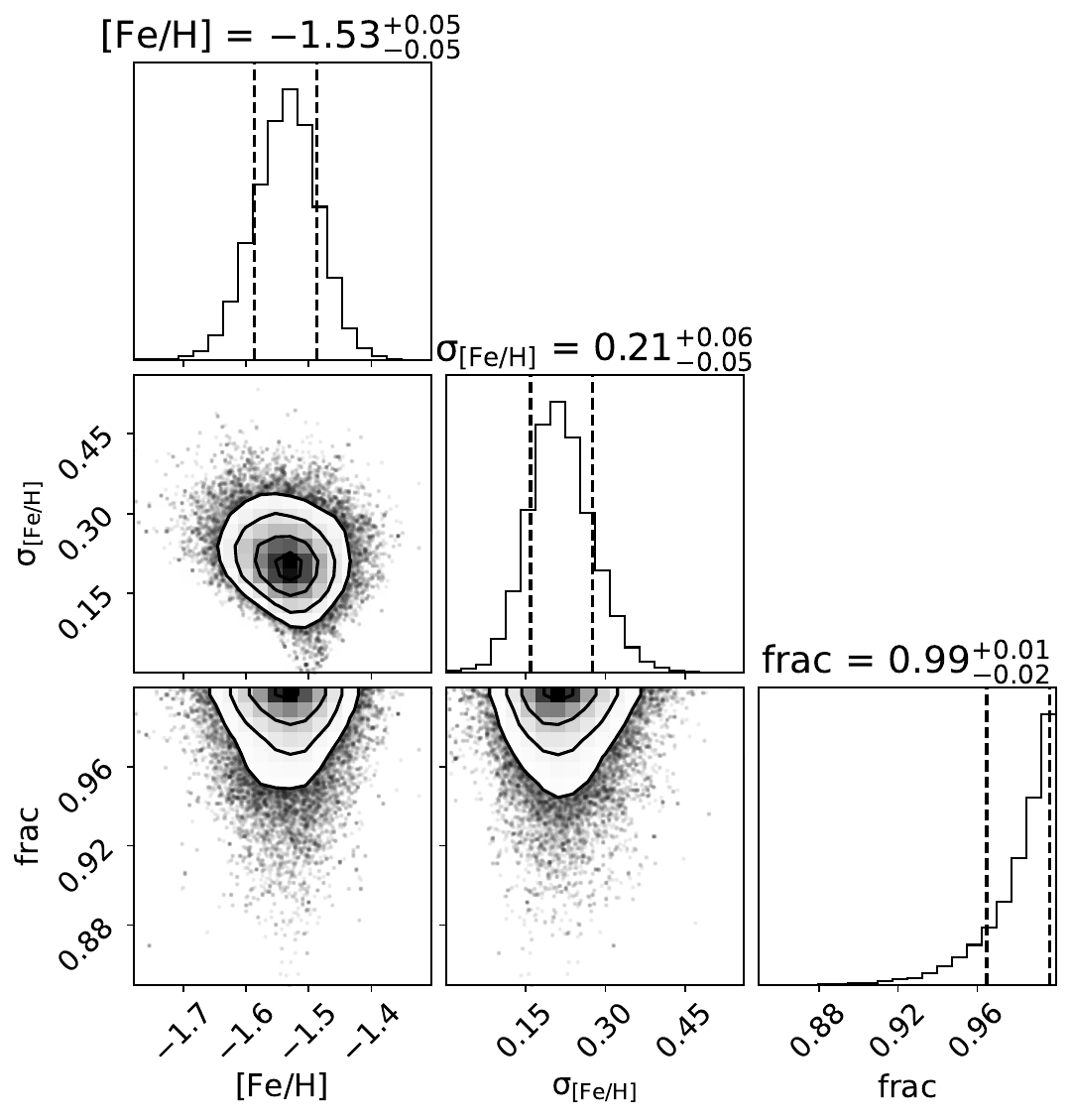}
        \caption{Corner plot for the MCMC metallicity fit using a
          sample of 55 stars. We show the mean metallicity, $\mu_u$,
          dispersion $\sigma_u$, and fraction of stars consistent with
          the model.} 
        \label{fig:mcmcmet}
\end{figure}

Of the 58 stars identified as Leo T members, \spexxy\ successfully
estimated the metallicity, [Fe/H], for 55 stars (it failed for the three emission line stars).
Figure \ref{fig:histfe} shows the histogram of the metallicity estimates. 
To quantify this distribution, we adopt the model described in Section \ref{sec:dist}. Therefore, we assume that
the underlying metallicity distribution is a Gaussian, which appears to be a reasonable approximation in this case. 
We note that stars in the wings of the distribution typically have low-S/N spectra. Furthermore, the stars with measurements of [Fe/H] $< -2.5$ have higher uncertainties on [Fe/H], as their spectra present almost no lines. We further explore the impact of these uncertainties in Appendix~\ref{sec:MetComp}. 

By fitting the model to the distribution, we find a mean value of $\mathrm{[Fe/H]} = -1.53 \pm 0.05$ and a dispersion of $\sigma_{\mathrm{[Fe/H]}}= 0.21 \pm 0.06$. The fit also considers that 99\% of the members of the sample are consistent with the model. The resulting corner plot of this MCMC fitting is shown in Figure~\ref{fig:mcmcmet}.

The results are consistent with those obtained from our photometric analyses, which is expected considering we used  [$\alpha$/Fe] = 0 in both measurements. However, as stated above, if we change the fixed value for [$\alpha$/Fe] in our fit with \spexxy\ we get different values accordingly.

To detect differences in metallicity between the young and old populations, we repeated the analysis separately for each population. Although the results appear to be consistent with each other (see Appendix~\ref{sec:MetComp}), we draw attention to the fact that the sample of the young population consists only of 17 stars, which is too small to put reliable constraints on the metallicity for the young population. 
Nevertheless, the constraints for the younger population are consistent with the results for the older population, although they favour a somewhat lower metallicity for the younger population. We discuss this further in Appendix~\ref{sec:MetComp}. 

\subsection{Stellar kinematics}\label{sec:kin}

Here, we start by discussing the overall kinematics for all the stars
with velocity measurements, combining MUSE and literature data.

\begin{figure}[!h]
\centering
\includegraphics[scale=.55]{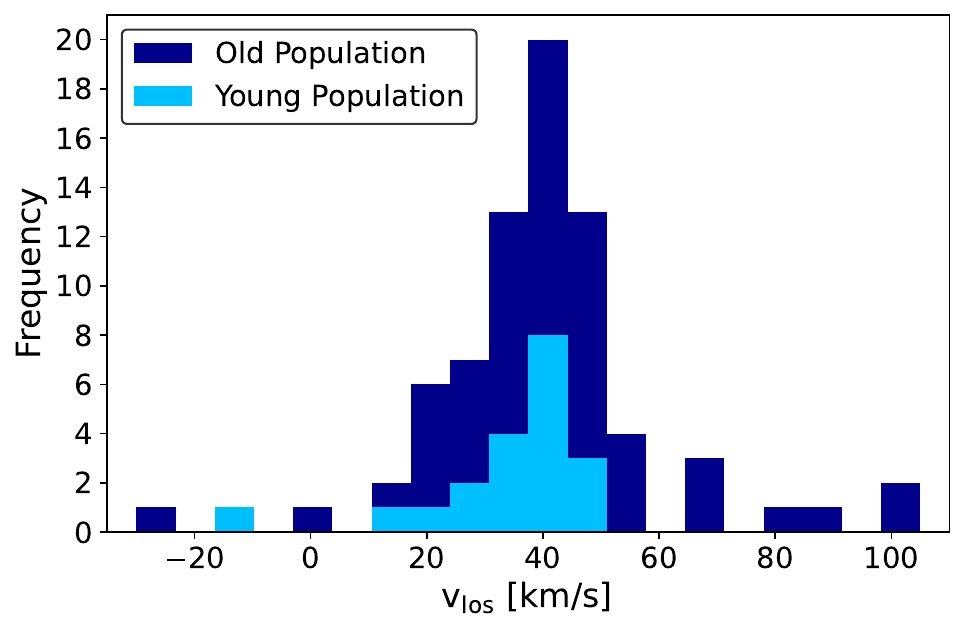}
\caption{Histogram of the line-of-sight velocities for Leo T members
  (75 stars), measured by fitting their spectra with \spexxy\ (55
  stars), \ULySS\ (3 stars), and adding the results of \cite{Simon_2007} (17 stars). The lighter colour corresponds to the younger population in the MUSE sample.}
        \label{fig:histv}
\end{figure}

\begin{figure}[!h]
\centering
\includegraphics[scale=0.45]{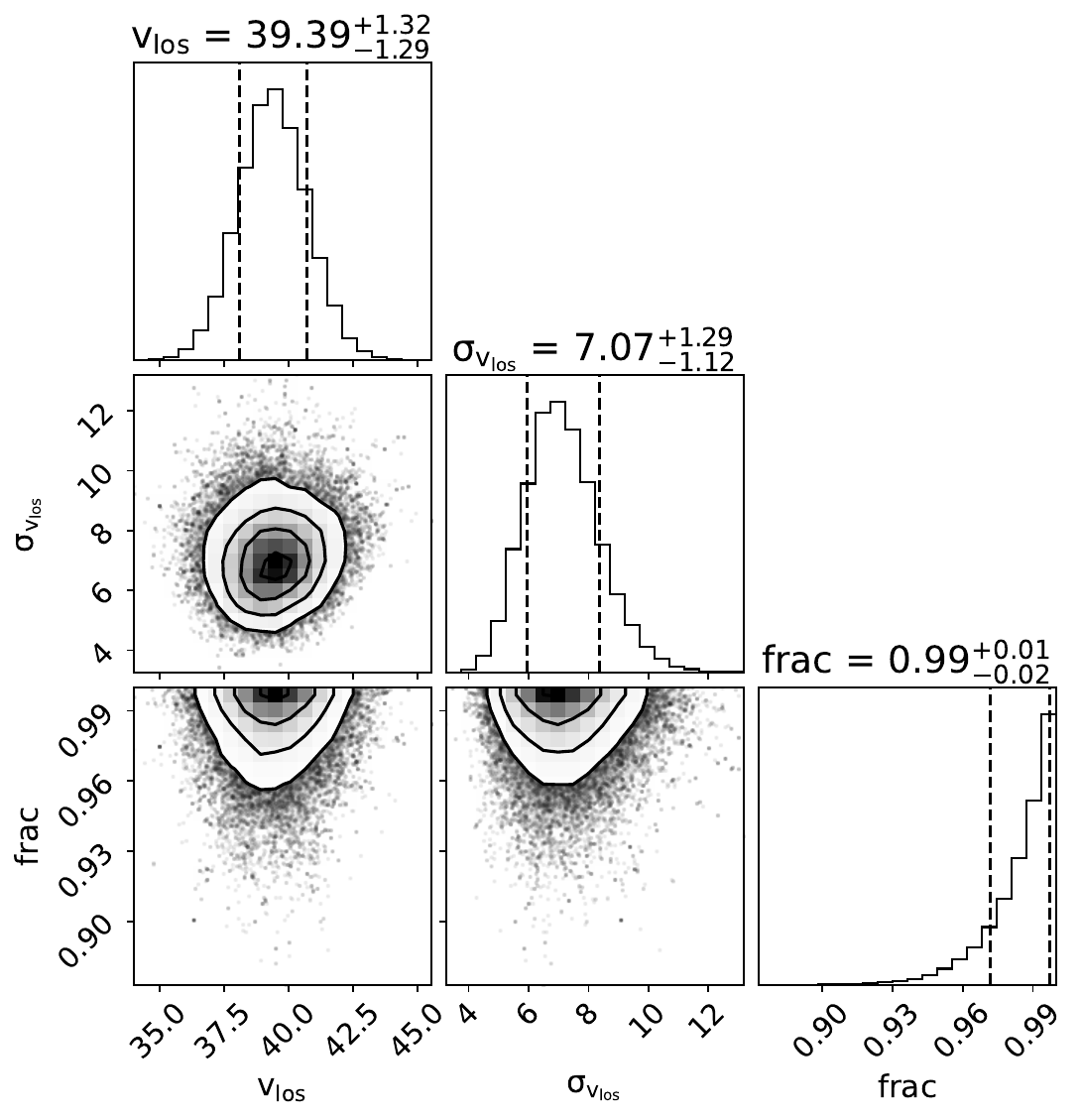}
\caption{Corner plot for the MCMC velocity fit using the entire
  sample of 75 stars. We show the mean value $v_{los}$, dispersion
  $\sigma_v$, and fraction of stars consistent with the model.}
        \label{fig:mcmcvel}
\end{figure}

As such, here we also used 17 of the 19 stars with kinematics measurements from \citet{Simon_2007}. Two stars were excluded because they overlap between the samples. Additionally, our line-of-sight measurements and those from the literature are consistent with each other for the two stars when considering the uncertainties. 

Therefore, the total sample for the kinematic analysis consists of 75 stars, which is the same sample presented by \citet{2021arXiv211209374Z}. The velocity histogram for these stars is displayed in Figure~\ref{fig:histv}.

To the data, we fit the model described in Section~\ref{sec:dist} using MCMC and the resulting corner plot is shown in
Figure~\ref{fig:mcmcvel}.  
We find a mean velocity of $v_{\mathrm{los}}=39.39^{+1.32}_{-1.29}\ \mathrm{km}\,\mathrm{s}^{-1}$, and an intrinsic velocity
dispersion of $\sigma_{v} = 7.07^{+1.29}_{-1.12}\ \mathrm{km}\,\mathrm{s}^{-1}$ with
99\% of the sample consistent with the model. 
The results are in agreement with both the measurements made by \cite{Simon_2007} and the \hi\ velocity and 
velocity dispersion measurements reported by~\citet{Adams2018}. 

We note here that, as the analysis deconvolves from the uncertainties on the velocities, it is important that the MUSE and \citet{Simon_2007} measurement
uncertainties are on the same scale. To verify this, we ran the MCMC with a slightly modified version of the model in Equation~\eqref{eq:likelihood} where we added a scale factor for the uncertainties of the \spexxy\ measurements. Marginalising this, we find that this scale factor is consistent with one, indicating that the measurement uncertainty estimates in the two datasets are consistent.

Similarly to what we did in Section~\ref{sec:met}, we explored any possible kinematical difference between the young and old stellar populations in Leo T.
We repeated the analysis with the sample divided into two sets: one with 20 stars that are consistent with ages of
$\leq 1\,\mathrm{Gyr}$ and another with 55 stars consistent with ages of $\geq 1\,\mathrm{Gyr}$. Here, we considered the entire sample of \citet{Simon_2007} to be composed of old stars.

\begin{figure}[!h]
\centering
\includegraphics[scale=0.45]{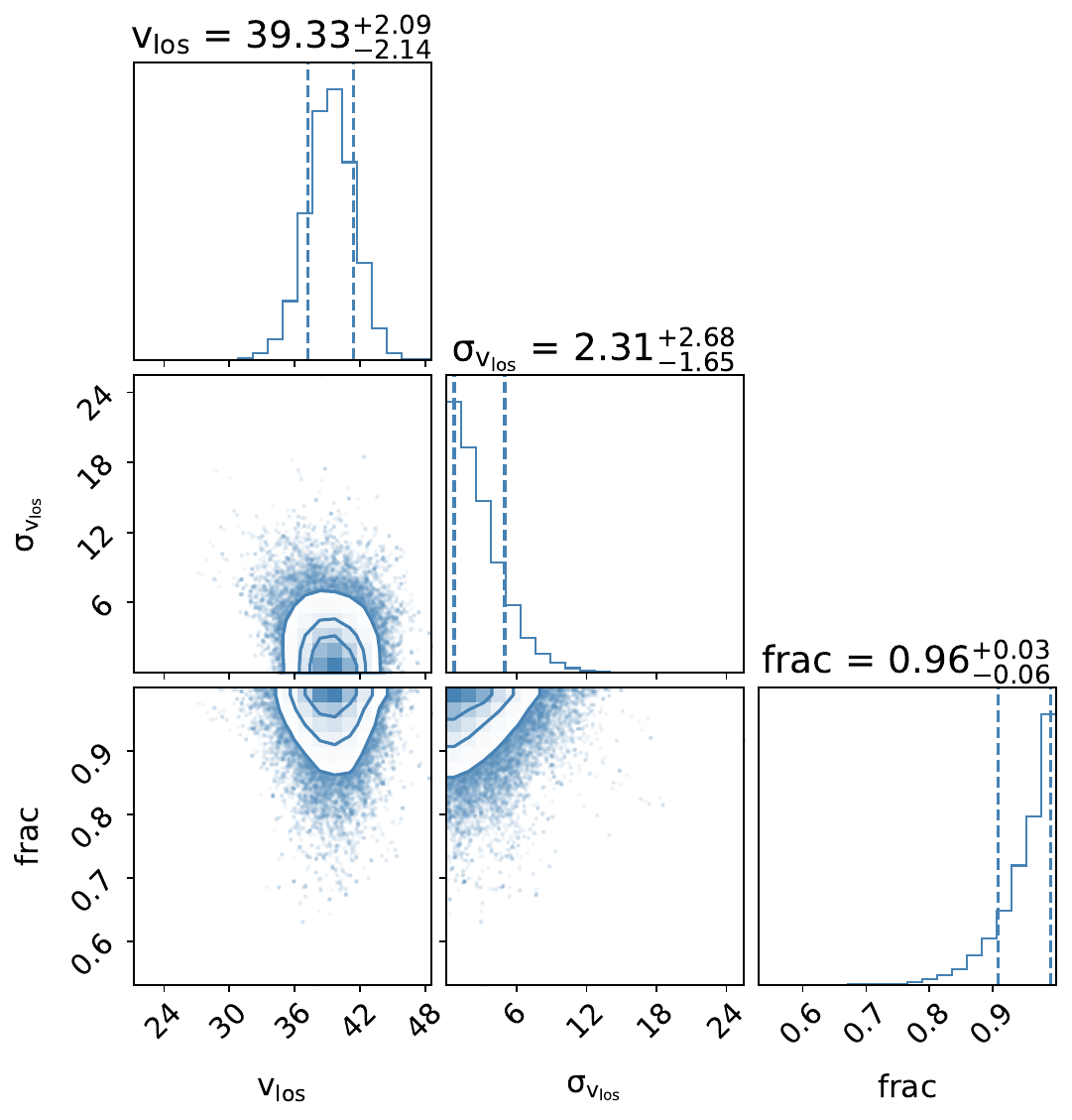}
\caption{Corner plot for the MCMC velocity fit using the sample of 20
  stars with ages of $\leq 1$ Gyr. We show the mean value
  $v_{\mathrm{los}}$, dispersion $\sigma_v$, and fraction of stars consistent with
  the model.}
        \label{fig:young}
\end{figure}

\begin{figure}[!h]
\centering
\includegraphics[scale=0.45]{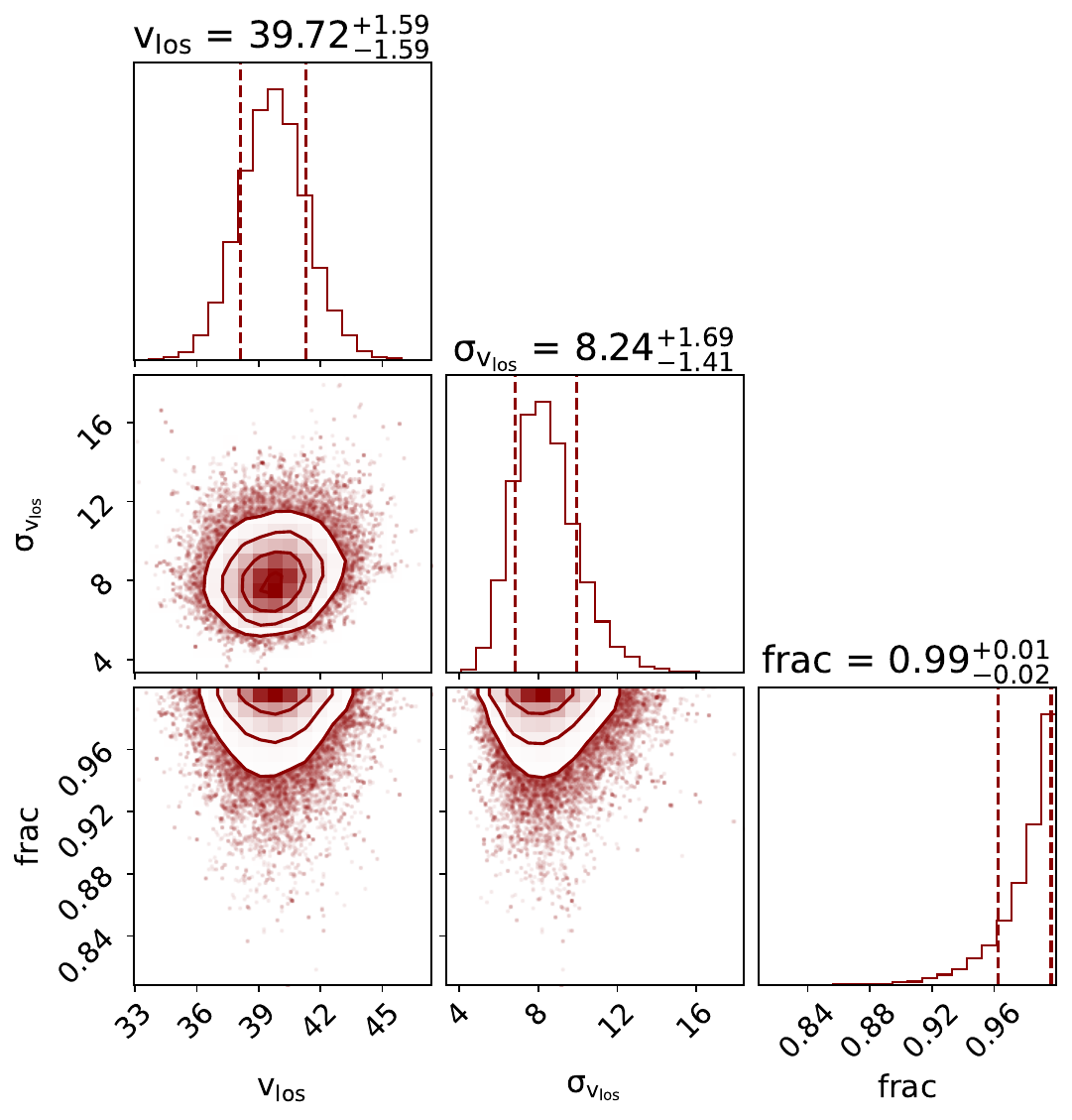}
\caption{Corner plot for the MCMC velocity fit using the sample of 55
  stars with ages of $\geq 1$ Gyr, including the data from
  \cite{Simon_2007}. We show the mean value $v_{los}$,
  dispersion $\sigma_v$, and fraction of stars consistent with the
  model.}
        \label{fig:old}
\end{figure}

Naturally, both sets were fitted with the MCMC method described above. This time, for the younger sample we get a mean value of
$v_{\mathrm{los}} = 39.33^{+2.09}_{-2.14}\ \mathrm{km}\,\mathrm{s}^{-1}$, a velocity dispersion
of $\sigma_{v} = 2.31^{+2.68}_{-1.65}\ \mathrm{km}\mathrm{s}^{-1}$, with 96\% of the sample being consistent with the model. For the older sample, we got a
mean value of $v_{los} = 39.72^{+1.59}_{-1.59}\ \mathrm{km}\,\mathrm{s}^{-1}$, a
velocity dispersion of
$\sigma_{v} = 8.24^{+1.69}_{-1.41}\ \mathrm{km}\,\mathrm{s}^{-1}$, with 99\% of the
sample being consistent with the model. The corner plots for these
results are shown in Figure~\ref{fig:young} for the young stars, and
Figure~\ref{fig:old} for the old stars.

\begin{figure}[h!]
\centering
\includegraphics[scale=0.45]{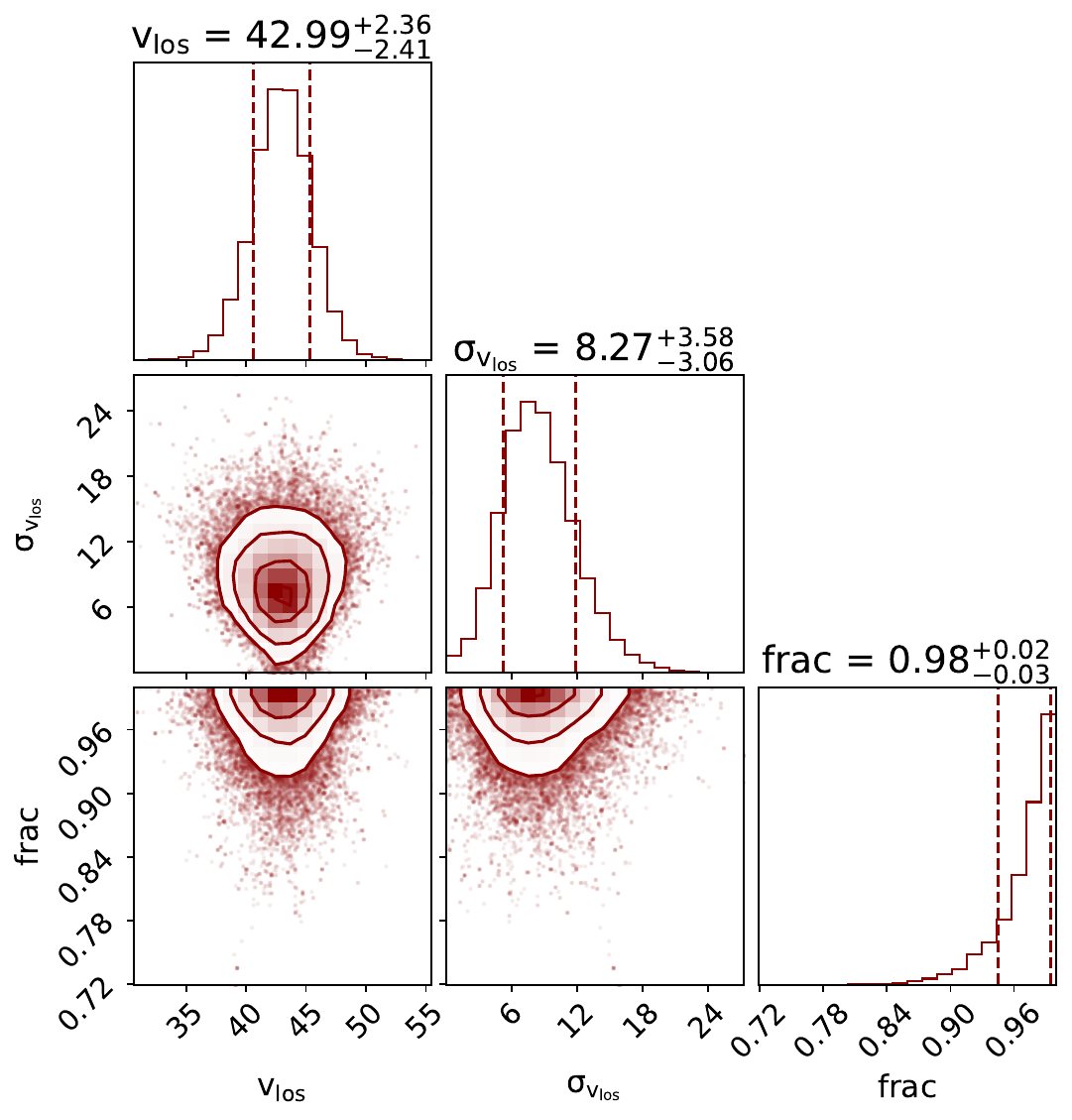}
\caption{ Corner plot for the MCMC velocity fit using the sample of 38
  stars with ages of $\geq 1$ Gyr within the MUSE FoV. We show the mean value $v_{los}$,
  dispersion $\sigma_v$, and fraction of stars consistent with the
  model.}
        \label{fig:oldmuse}
\end{figure}

 As only the set of older stars includes data from \cite{Simon_2007}, we also repeated the analysis for the older stars but considering only the 38 MUSE stars. The estimated intrinsic velocity dispersion is almost the same as before, but the estimated mean velocity increases by $\sim~3~\mathrm{km}\,\mathrm{s}^{-1}$.
 The corner plot for these results is shown in Figure~\ref{fig:oldmuse}. Either way, the younger stars show a significantly smaller velocity dispersion
than the older stars.

\begin{figure}[]
\centering
\includegraphics[scale=0.5]{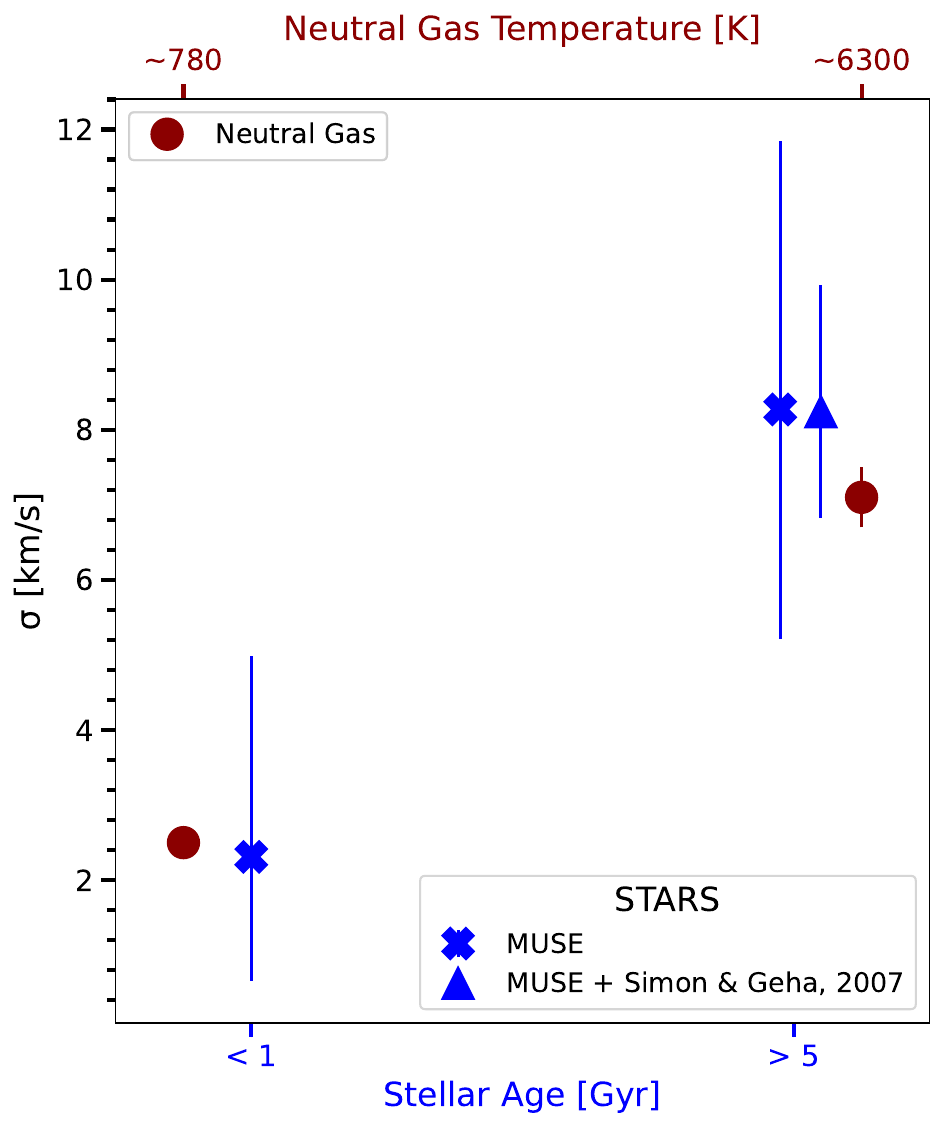}
\caption{ Comparison between stellar velocity dispersion of this work and neutral gas velocity dispersion from \cite{Adams2018}. 
For the older stellar population, we present two values: estimation using only the MUSE sample and estimation using MUSE + literature sample from \cite{Simon_2007}}
        \label{fig:disp_comp}
\end{figure}

\begin{figure}
    \centering
    \includegraphics[scale=0.5]{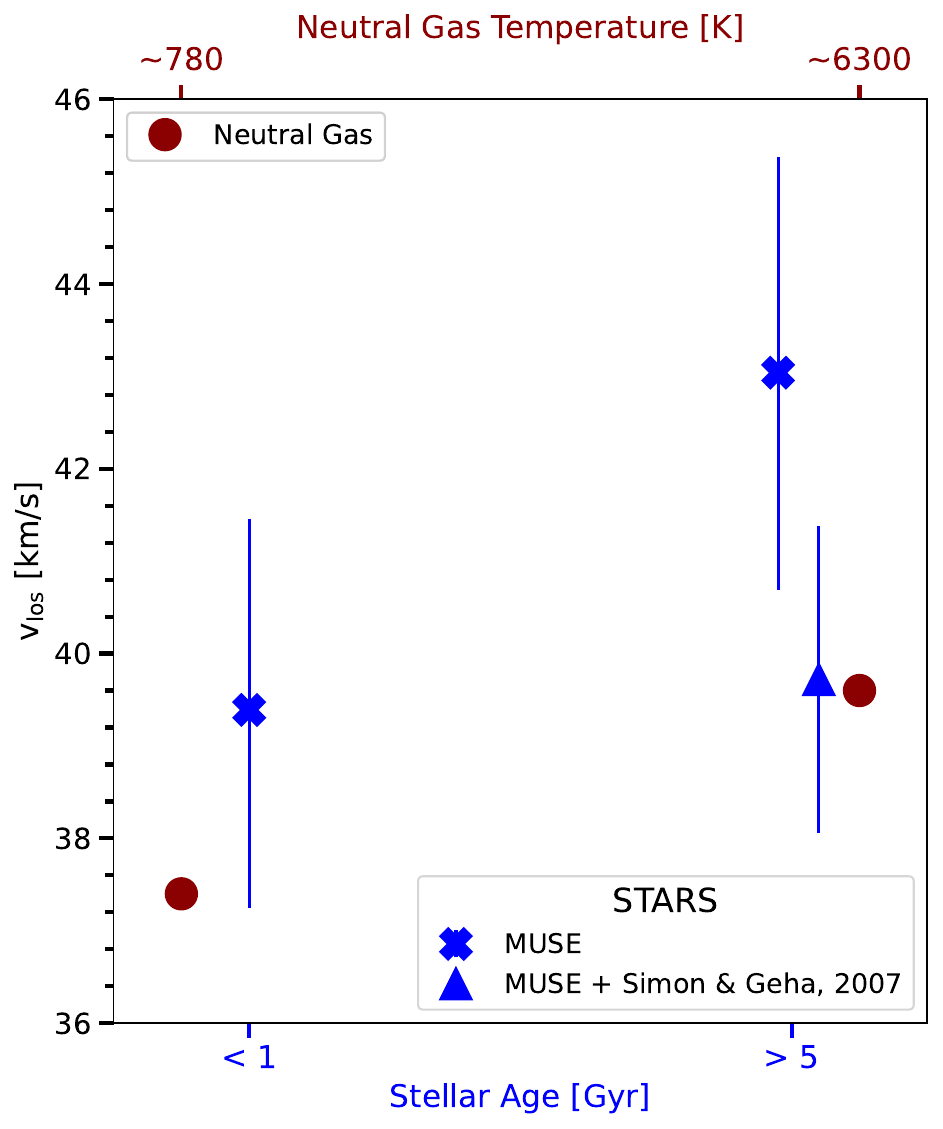}
    \caption{Comparison between stellar velocity of this work and neutral gas velocity from \cite{Adams2018}. For the older stellar population, we present two values: estimation using only the MUSE sample and estimation using MUSE + literature sample from \cite{Simon_2007}}
        \label{fig:vel_comp}
\end{figure}

We now compare the differences in kinematics between young and old stars with what was found by \cite{Adams2018} when analysing the \hi\ kinematics of warm and cold neutral gas (see also Table~\ref{tab:LeoTP}). 
This comparison can be seen in Figures~\ref{fig:disp_comp} and \ref{fig:vel_comp}, where we show the measured values stated above and the measurements of \cite{Adams2018} that are displayed in Table~\ref{tab:LeoTP}.

We find a good match when comparing the velocity dispersion of the young population with the cold component of the \hi\ gas, and between the kinematics of old Leo T stars and the warm component of the \hi\ gas, as shown in Figure~\ref{fig:disp_comp}.
The most natural explanation for this is that the current cold \hi\ gas is representative of the gas from which this young population of stars
formed and that these stars have not yet been heated dynamically to the velocity dispersion of the halo. 
On the other hand, assuming that the warm \hi\ gas has remained in the system for a significant time, both the WNM and the old stars are accurately tracking the gravitational potential of the galaxy.

We also compare the mean velocity of both populations of stars with both components of the neutral gas (see Figure~\ref{fig:vel_comp}). Here, and using the entire sample, we find negligible differences between the mean velocity of the young and old populations, which contrasts with the difference in the mean velocity found for the cold and warm components of the \hi\ gas.

Nevertheless, this is not the case when the analysis is performed using only the MUSE stars. In that case, we find a difference of $\sim~3~\mathrm{km\ s^{-1}}$ between the mean velocity of old and young stars. 
We are unable to identify the reason for the offset. We note that the telluric correction applied by \cite{Simon_2007} is different from the way our velocity estimates are made. Also, the spatial distribution of the stars in the \cite{Simon_2007} sample is different from that of the MUSE sample, as MUSE observations target the innermost stars of Leo T. 

\subsection{No extended emission line sources}
\label{sec:no_emline_sources}

The presence of young stars raises the question of whether there is ongoing star formation in Leo T. The MUSE data are particularly well-suited for this as they allow for the construction of very deep narrow-band images over lines of interest.

\begin{figure}
  \centering
  \includegraphics[width=0.5\textwidth]{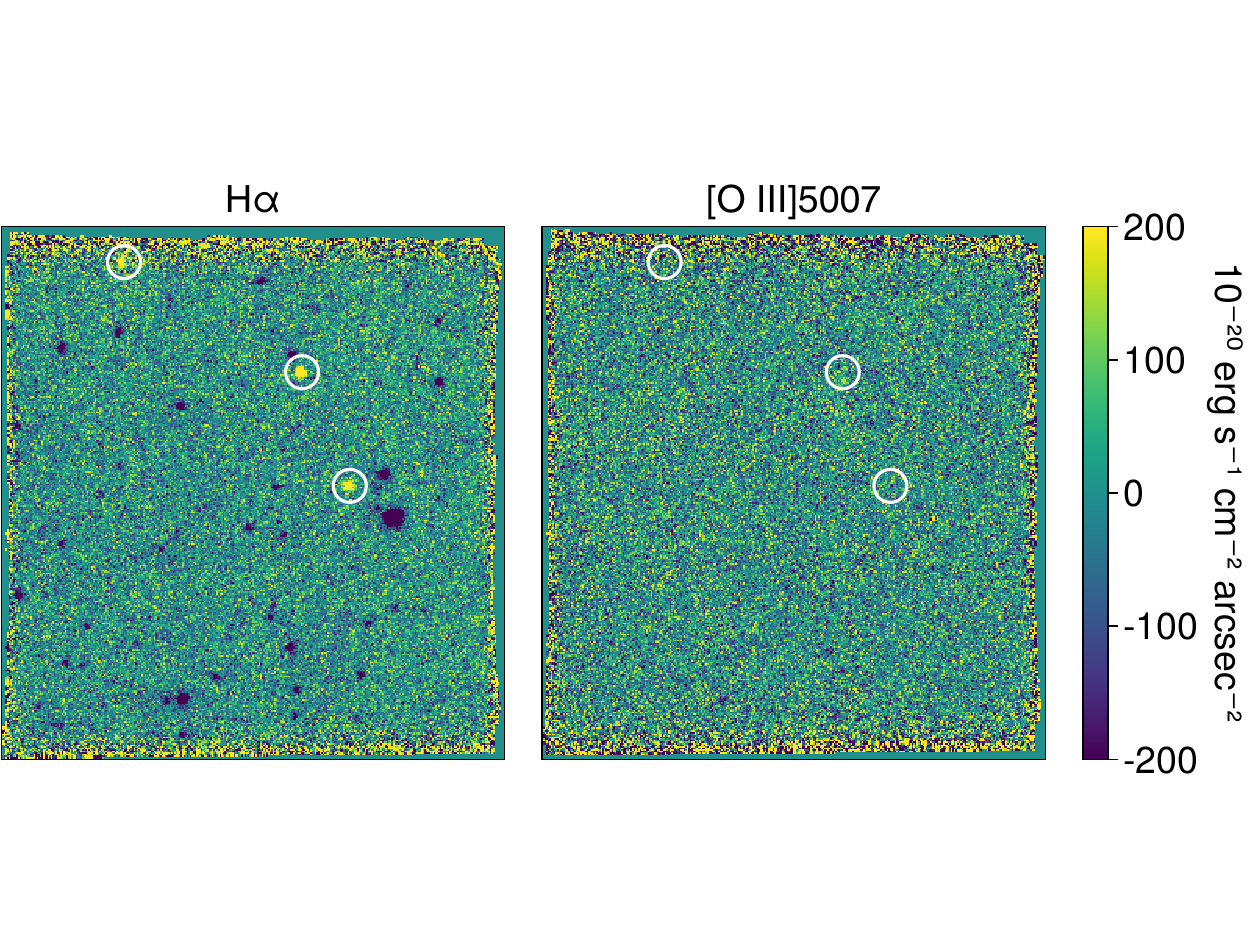}
  \caption{Narrow-band images over \ha\ (left) and \oiii{5007}
    (right). In the left-hand panel the three emission line stars are
    easily seen, but no evidence is found of extended emission in either
    \ha\ or \oiii{5007}.}
  \label{fig:nb_images}
\end{figure}

Figure~\ref{fig:nb_images} shows the narrow-band images over \ha\ and \oiii{5007}. These were created by weighting a Gaussian
centred on the wavelength of the lines adjusted by the line-of-sight velocity of Leo T, and with a velocity width of $\sigma=70~\mathrm{km~s^{-1}}$. 
We sum this up over~$\pm 5\sigma$ after removing the background. We estimate the background as the median over $\pm 100$\AA\, excluding the
$\pm 5\sigma$ region.

The \ha\ image shows a number of dark sources. These are stars in the FoV showing \ha\ absorption close to the velocity of Leo T. The three
bright sources circled in white are the three emission line stars discussed above.

There is no sign of extended emission in either image. To estimate a detection limit, we calculated an empirical noise estimate by randomly
distributing 100 apertures over the narrow-band images and calculating the median absolute deviation of the measured median level. This leads
to a $1 \sigma$ limit for \ha\ of $0.99\times 10^{-20}$ and $24.89\times10^{-20}$ per pixel and per squared arcsecond, respectively, in units of
$\mathrm{erg~s^{-1}~cm^{-2}}$ for a single line with line width $70~\mathrm{km~s^{-1}}$. 
For \oiii{5007,} the values are $1.05\times 10^{-20}$ and $26.29\times10^{-20}$ respectively. 

We repeated the calculation using a data cube not corrected with ZAP to check for differences. We get a $1 \sigma$  limit for \ha\ of $1.04\times 10^{-20}$ and $25.92\times10^{-20}$ per pixel and per arcsecond squared, respectively, in units of $\mathrm{erg~s^{-1}~cm^{-2}}$ for a single line with line width $70~\mathrm{km~s^{-1}}$. For \oiii{5007,} the values are $1.11\times 10^{-20}$ and $27.73\times 10^{-20}$ respectively. 
Therefore, we conclude that the use of ZAP to remove residual sky signatures does not affect our results here.

We convert the \ha\ flux into an upper limit on star formation rate (SFR) using the $\mathrm{L}_{\ha}$--SFR relation derived by \cite{Kennicutt_1998}.
We get a SFR~$\approx 4 \times 10^{-11}~\mathrm{M_\odot~yr^{-1}}$ per arcsecond, corresponding to a SFR~$\sim 10^{-11}~\mathrm{M_\odot~yr^{-1}~pc^{-2}}$.
It is worth noting that the SFR calibration assumes solar metallicity and a fully populated initial mass function (IMF), neither of which are likely to be satisfied in Leo T. However, even allowing for this, the SFR upper limit is extremely low, which is an indication of the lack of ongoing star formation in Leo T.

\section{Discussion}
\label{sec:discussion}

We find that Leo T has two stellar populations that have significantly different age and dynamics when compared to the neutral gas. However, the metallicity of the two populations appears to be nearly identical. We also see that the young population, despite its low metallicity, contains emission-line Be stars with properties comparable to those found in surveys of more metal-rich populations \citep{porterClassicalBeStars2003}. Despite the presence of young stars, we do not detect extended emission-line sources.
Here, we discuss the results and compare them with those from the literature.

\subsection{Metallicity and star formation}

The most immediate interpretation of the lack of extended emission line sources is that star formation has ended. \citet{Weisz_2012}, based on the stellar photometry, argue that star formation ended about 25 Myr ago, which is consistent with the lack of extended emission from any \hii\ region. Therefore, our observations corroborate the picture presented by~\citet{Weisz_2012} of essentially no star formation in Leo T in the last $\sim 25$ Myr. This is consistent with recent models that suggest that star formation in low-mass galaxies should be bursty with short quiescent periods \citep[e.g.][]{Collins_2022}.

This is also consistent with the low dispersion in [Fe/H] in Leo T, which implies that all stars have approximately the same metallicity of [Fe/H] $\sim - 1.6$. This can be compared with the findings of \cite{Weisz_2012}. We do see similar behaviour in other dwarf galaxies (e.g. Aquarius \citep{Cole_2014}, and theoretical models have predicted that metals can escape from low-mass dwarfs due to stellar feedback, keeping [Fe/H] approximately constant \citep{Jeon_2017, Emerick_2018}, which is also enabled by a bursty star formation.
Nevertheless, this appears to happen only after an initial chemical evolution, which we do not see in Leo T. We measure the same metallicity for stars consistent with ages as low as $\sim 200$ Myr and as high as $\sim 10$ Gyr, which is rather surprising given the interpretation of Leo T as a galaxy with near-constant star formation over cosmic time.

Additionally, a value of [Fe/H] $\sim - 1.6$ makes the older stars of Leo T quite metal-rich compared to what we see in other dwarfs \citep{Jeon_2017}. 
As we cannot confirm the age of the stars, due to the degeneracy between the different isochrones, there is the possibility that these stars are younger than is indicated by our fit. We could be missing an older and more metal-poor population in our analysis, which could also explain why our measured metallicity is not identical to the value of \cite{Simon_2019}, assuming that this difference does not result from differences in calibration or data processing.
However, there is no clear reason as to why we would not detect such population, considering that we are observing the Leo T field. 
It is worth noting that a significant portion of our sample has been removed due to low S/N. Typically, these stars are of low luminosity, which may indicate that they are quite old and, presumably, have low metal content. Therefore, it is possible that the members of the population we assume is missing could be among the stars that were rejected due to low S/N.
To explore this possibility, we look at the sample of 194 spectra that were not used in our analysis due to the S/N cut or because of a failed \spexxy\ fit. 
Naturally, these spectra share a commonality in that they are very faint. In the colour--magnitude space, they generally fall between F814W~$\sim 24$ and $26$ and between $\mathrm{F606W-F814W}~\sim 0$ and $0.6,$ which suggests the presence of multiple age and/or metallicity populations. Nonetheless, due to the faintness of those objects (and respective low S/N) we are unable to fit their spectra and verify the hypotheses that an older and more metal-poor population exists. We draw attention to the fact that these results are in line with the observations of \cite{Weisz_2012}.

Subsequently, we cannot discard the possibility that the older Leo T stars present a value of [Fe/H] $\sim - 1.6,$ which would indicate that Leo T was chemically enriched before starting to produce stars. Deeper observations are needed to solve this dilemma.

\subsection{Recent star formation and link to the CNM}

In~\citet{Adams2018}, the fact that Leo T has a significant amount of its \hi\ mass in the CNM
phase but a low star-formation efficiency compared to other dwarf galaxies was discussed in some detail.
The authors highlighted a number of possible explanations, but emphasised in particular the possibility that the formation of the CNM is recent and connected to the infall of Leo T into the Milky Way halo and is therefore less connected to the star formation in Leo T.

However, here we find that the CNM and the young stars in Leo T have comparable dynamical properties, and therefore a close link between the two appears natural. Both have significantly smaller velocity dispersions than the WNM and the older stars. 
In this context, it is natural to ask how long it would take for the young stars or the CNM to
approach the dynamical state of the older stars.

 We can estimate the relaxation time of the system, if we write it as \citep{spitzerRandomGravitationalEncounters1971}
\begin{equation}
  \label{eq:t_relax}
  t_{\mathrm{relax}} = \frac{v^2}{D(\Delta v^2)}, 
\end{equation}
 where $D(\Delta v^2)$ is the diffusion coefficient and has
  the greatest impact, which for a spherical system and an isotropic
  Maxwellian velocity distribution for both young and old stars can be
  written as
\begin{equation}
  \label{eq:Diffcoeff}
 D(\Delta v^2) = \frac{4\sqrt{2} \pi G^2 \rho m_a \ln \Lambda}{\sigma} \frac{G(X)}{X},
\end{equation}
where $m_a$ is the mass of the younger stars and $\sigma$ their
velocity dispersion, $\ln \Lambda$ is the Coulomb logarithm and
$X=v/\sqrt{2}\sigma$, with $v$ denoting the typical velocity of the
older stellar component. We approximate the Coulomb logarithm as
\begin{equation}
  \label{eq:CoulombLogarithm}
  \ln \Lambda = \frac{R\sigma_{\mathrm{old}}^2}{G
    \left(m_{\mathrm{old}}+m_{\mathrm{young}}\right)},
\end{equation}
where we take $R$ to be the half-light radius of Leo T (see
Table~\ref{tab:LeoTP}) and $\sigma_{\mathrm{old}}$ to be the velocity
dispersion of the old component. The IMF-weighted average mass of the
old and young population will differ, but
the difference is modest  for the ages relevant here, and so for order-of-magnitude estimates we can
set them equal to 0.3 M$_\odot$. 

Taking the velocity dispersions given in Section~\ref{sec:kin}, and
the density profile and mass-to-light ratio presented
in~\citet{2021arXiv211209374Z}, we estimate relaxation times
of $>10^{13}\,\mathrm{yr}$. While the relaxation time could be reduced if
dark matter is made up of massive compact halo objects (MACHOs), these latter would need to have masses that are excluded by other
constraints
\citep[e.g.][]{brandtConstraintsMACHODark2016,Zoutendijk_2020} in
order to bring the relaxation time below a Hubble time.

Therefore, we can conclude that the kinematics of the young population
is likely a fair reflection of the kinematics of the gas it formed out
of. As the young stellar population shows very similar kinematics
to the CNM, the natural inference is that the most recent star
formation in Leo T originated from gas whose remnants now make up the
CNM. However, this also means that the overall velocity dispersion that we find above, which includes contributions from the young population whose kinematics may not yet be fully relaxed, may not reflect the true gravitational potential of Leo T. Therefore, the velocity dispersion of the older stars may provide a more reliable probe of the gravitational potential of the system.

The different kinematics of the CNM and WNM, and more precisely the different mean velocity, might
imply an external origin for the CNM, through accretion for example. 
Another result that would support the extragalactic accretion argument is that of the metallicity of the younger population of Leo T.
The fact that the younger population could have a lower metallicity than the older stars might imply that the younger stars formed from gas less chemically enriched than that of the gas from which the older stars formed; an external origin for this gas would explain this observation.

Another natural explanation is related to the infall of Leo T into the Milky Way circumgalactic medium. Indeed, the ram pressure affects the hot, more extended gas more rapidly than the cool gas \citep{Emerick_2016}, and this could explain the difference in kinematics between the two components of the gas. This would lead to changes in the internal structure of the gas, which would also explain the recent quench in star formation in Leo T.

\subsection{Be stars in metal-poor environments}
Turning now to the properties of young stars, in particular the
emission-line Be stars identified in Section~\ref{sec:emstars}, their
presence offers some insight into the formation of stars at low
metallicity. We find that 15\% of the stars classified as young show
\ha\ in emission, which can be compared to the Milky Way study of
\citet{Mathew_2008}, who reported rates at a level of
10\%--20\% in stellar clusters. \citet{Schootemeijer_2022}, who present a study of Be
stars in metal-poor environments down to one-tenth of solar, find OBe
fractions of between $0.2$ and $0.3$ when limited to main sequence stars. This is consistent with what is found in Leo A, where \cite{Gull_2022} recently reported the spectroscopic detection of Be stars at a metallicity of one-tenth of solar. Meanwhile, \cite{Alina_2022} found that, for Leo A, the emission-line stars account for up to $\sim$15\% of the blue helium-burning (BHeB) stars.
Similar values are also found in other metal-poor environments, such
as NGC 2345 with $[\mathrm{Fe/H}] =~ - 0.28$
\citep{alonso-santiagoComprehensiveStudyNGC2019}.

As the presence of emission lines in these stars is thought to be
linked to high rotational velocities, these findings suggest that extremely
rapidly rotating massive stars are common in low-metallicity
environments \citep{Schootemeijer_2022}. Leo T pushes this down to $1/30$ solar metallicity, and
although the number of stars is too small to place strong constraints,
the fraction of emission line stars on the main sequence in Leo T is
high. Coupled with the large velocity width seen in the \ha\ line,
this supports the conclusions
of~\citet{Schootemeijer_2022}. 

\section{Conclusions}
\label{sec:conclusions}

We present a MUSE study of Leo T, one of the galaxies with the  lowest mass showing signs of recent star formation. 
We were able to increase the number of spectroscopically observed Leo T stars from 19 to 75, which led to the following results.

By combining the MUSE data with HST photometry and performing a photometric analysis, we identified two stellar populations ---in agreement with previous results in the literature--- with a wide range of stellar ages, from as low as $200$~Myr to as high as $10$~Gyr. 
We divide the population into two samples such that the younger population shows consistency with ages of $<500~\mathrm{Myr}$, while the older population shows ages of $>5~\mathrm{Gyr}$. Within the younger population, we identified three emission-line Be stars,
corresponding to 15\% of the stars classified as young. This is in line with results for the Milky Way, although these stars show much lower metallicities, and our findings therefore suggest a substantial number of rapidly rotating young stars at 1/30 solar metallicity.

The presence of young stars could indicate that there is ongoing star formation in Leo T, but we find no evidence of the existence of extended emission-line
regions to a surface brightness of $< 1\times 10^{-20}\,\mathrm{erg}\,\mathrm{s}^{-1}\,\mathrm{cm}^{-2}\,\mathrm{arcsec}^{-2}$, which corresponds to an upper limit SFR~$\sim 10^{-11}~\mathrm{M_\odot~yr^{-1}~pc^{-2}}$.
Therefore, we conclude that star formation has ended, which is consistent with \citet{Weisz_2012}, who conclude that there has been no star formation in the last $\sim 25$ Myr. 

By fitting the MUSE spectra, we obtained a metallicity of $\mathrm{[Fe/H]} = -1.53 \pm 0.05$, which is in good agreement with our photometric analysis, assuming that [$\alpha$/Fe] = 0. We also find a metallicity dispersion of $\sigma_{\mathrm{[Fe/H]}}= 0.21 \pm 0.06$. The low dispersion in [Fe/H] shows that all stars have very similar metallicity, implying that Leo T underwent almost no evolution throughout its history, which is puzzling considering the claimed extended star formation history and age range (from $200$ Myr up to $10$ Gyr) of this dwarf. This seems to indicate that a large fraction of metals were ejected, namely due to stellar feedback. This has been predicted by theoretical models that also show that star formation in low-mass galaxies should be bursty with short quiescent periods, and these findings support that conclusion.

Using the sample of 75 stars, we measured the kinematics of Leo T stars in the most robust
kinematic study of this dwarf to date. We find a velocity dispersion of $\sigma_{v} = 7.07^{+1.29}_{-1.12}\ \mathrm{km}\,\mathrm{s}^{-1}$. This result is consistent with previous estimates of the kinematics of stars, but, more importantly here, it is also concordant with the measured \hi\ gas kinematics. Interestingly, when splitting the sample into young and old stars, we find that they have different kinematics. We find a velocity dispersion
of $\sigma_{v} = 2.31^{+2.68}_{-1.65}\ \mathrm{km}\,\mathrm{s}^{-1}$ for the younger population and a much higher velocity dispersion of $\sigma_{v} = 8.24^{+1.69}_{-1.41}\ \mathrm{km}\,\mathrm{s}^{-1}$ for the older population. This means that the cold component of the \hi\ gas and the young stars in Leo T have comparable kinematics. 

We estimate the time that it would take for young stars or the cold component of \hi\ gas to approach the dynamical state of older stars to be higher than the Hubble time, which directly links the most recent star formation in Leo T to the cold component of the \hi\ gas.

\begin{acknowledgements}

DV and JB acknowledge support by Fundação para a Ciência e a Tecnologia (FCT) through the research grants
UID/FIS/04434/2019, UIDB/04434/2020, UIDP/04434/2020. DV acknowledges support from the Fundação para
a Ciência e a Tecnologia (FCT) through the Fellowship 2022.13277.BD.
JB acknowledges financial support from the Fundação para a Ciência e a
Tecnologia (FCT) through national funds PTDC/FIS-AST/4862/2020 and
work contract 2020.03379.CEECIND. SLZ acknowledges support by The Netherlands Organisation for Scientific Research (NWO) through a TOP Grant Module 1 under project number 614.001.652. SK acknowledges funding from UKRI in the form of a Future Leaders Fellowship (grant no. MR/T022868/1). 
PMW acknowledges support by the BMBF from the ErUM program (project VLT-BlueMUSE, grant 05A20BAB).
Based on observations made with ESO Telescopes at the
La Silla Paranal Observatory under programme IDs  0100.D-0807, 0101.D-0300,
0102.D-0372 and 0103.D-0705. This research has made use of Astropy \citep{astropy, astropy2018}, corner.py \citep{FM2013}, matplotlib \citep{Hunter:2007}, NASA’s Astrophysics Data System Bibliographic Services, NumPy \citep{harris2020array}, SciPy \citep{2020SciPy-NMeth}.

\end{acknowledgements}

%
%
\bibliographystyle{aa}
\bibliography{biblio}

\begin{appendix}

\section{\textit{ULySS}}\label{sec:uly}

To check the robustness of the velocity
estimation method we used in this work, we follow the approach of \cite{Roth_2018}, using
\ULySS\ \citep{Cappellari04, Koleva_2009} and the empirical
\texttt{MIUSCAT} library \citep{Vazdekis_2012} to fit the stellar
spectra, as an alternative method to measure the radial velocity. 
We did this in two different ways: by fitting a single stellar
spectrum from the library to the full spectrum and using this to get
$v_{\mathrm{los}}$; and by fitting multiple reference spectra to the observed spectrum and using this composite spectrum to get
$v_{\mathrm{los}}$. The composite spectrum is created by combining different spectra from the library in order to achieve a best fit. This method could in particular be useful
for identifying observed spectra that are a composite of the spectra of multiple objects, e.g. from binaries.

As such, we fit each individual spectrum twice, and we consider the result corresponding to the best fit. We compare the results of \ULySS\ and \spexxy\  for the sample of 55
stars (as \spexxy\ failed to fit the emission-line star spectra) and look for discrepancies.

\begin{figure*}[hbt!]
    \centering
    {\includegraphics[scale=.8]{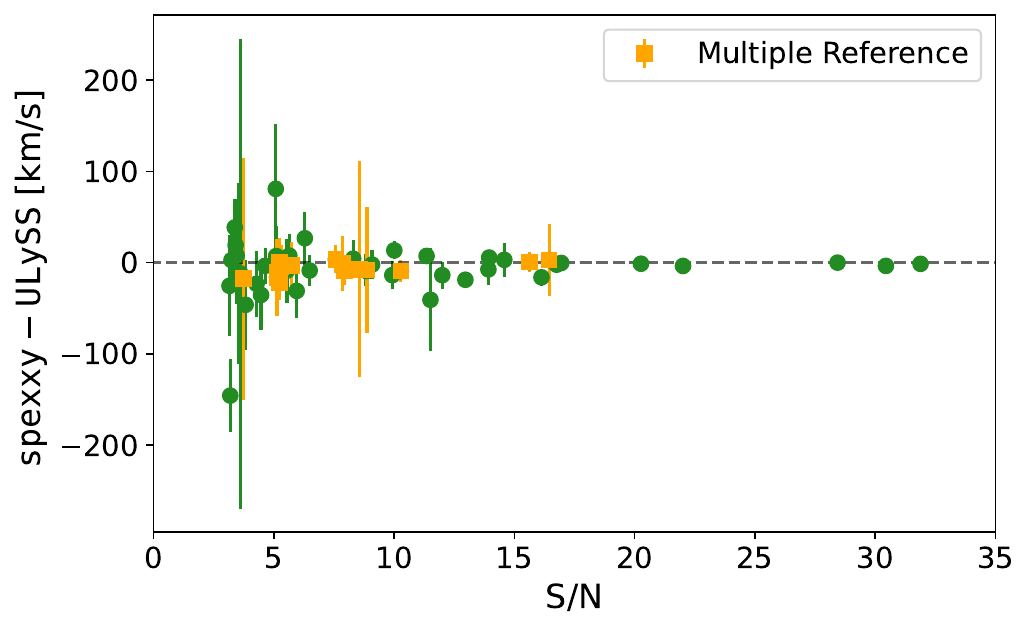}}
    \caption{Difference between \spexxy\  and \ULySS\ velocity measurements as a function of  S/N. The displayed error was determined by summing the respective measurement uncertainties in quadrature. The stars for which the \ULySS\ best-fit run used multiple reference spectra are
displayed as orange squares. A horizontal dashed line at zero is also shown.}
        \label{fig:diffit}
\end{figure*}

Figure~\ref{fig:diffit}  shows the difference in the radial velocity
estimates between \spexxy\  and \ULySS\ against the measured S/N. The displayed error is determined by summing in quadrature the respective measurement uncertainties. The stars which \ULySS\ best fit employ reference spectra of multiple spectral types (composite spectra) are shown in orange.
From the results, it is clear that both measurements are consistent
with each other within the errors. For a higher S/N, the difference in both measurements is effectively zero.
We do see a larger dispersion for lower S/N values, which is to be expected, but this is compensated for by the larger uncertainties.

Therefore, we concluded that the two velocity estimation
methods give consistent results and adopted \spexxy\ as the main method to 
fit the spectra in the analysis. As the \spexxy\ runs failed for the emission
line spectra, we used the \ULySS\ measurements for these objects in the kinematic analysis.

We note that \ULySS\ favoured a complex spectrum fit for 15
stars, which might indicate unresolved binaries, but we mention here
that the final spectra are averages over several independent epochs, and therefore they are likely to be less affected by binary motions than
single-epoch spectra, and we do not attempt to correct for this here.

 \FloatBarrier
\FloatBarrier

\section{[Fe/H] Analysis complement }\label{sec:MetComp}

\begin{figure*}[]
\centering
 {\includegraphics[scale=.6]{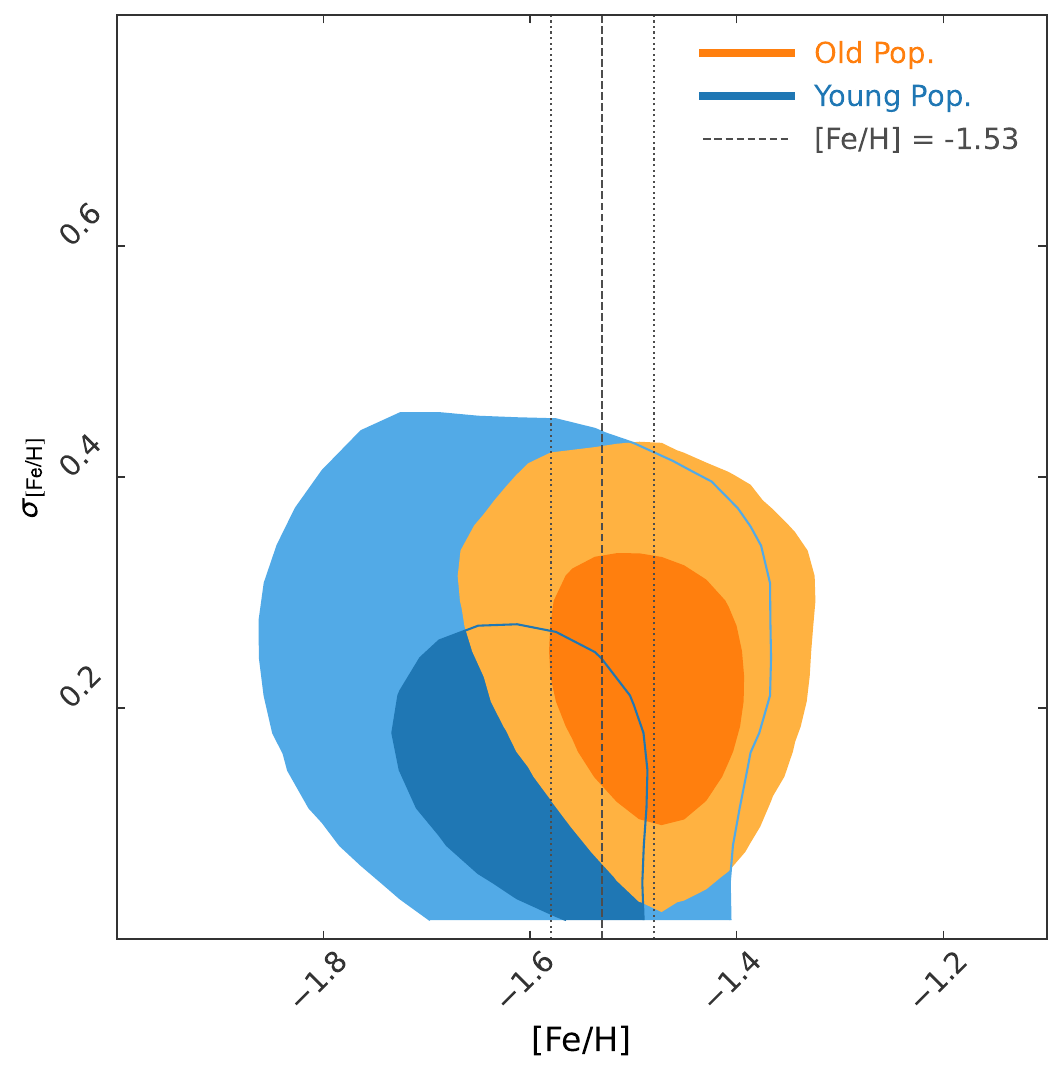}}
\caption{Plot of the covariance distribution of the mean [Fe/H] and [Fe/H] dispersion drawn from the MCMC analysis. The result for the young and old population is plotted in blue and orange, respectively. The darker shade symbolises the middle 68\% of the distribution. The dashed lines represent the mean [Fe/H] and the uncertainty obtained by fitting the entire sample with the same MCMC method.}
        \label{fig:feh_young_vs_old_cov}
\end{figure*}

Here we discuss the metallicity analysis when performed separately for the young and old population. From the sample of 55 stars with [Fe/H] measured spectroscopically, 17 are members of the young population. For these two samples, we fit the MCMC model described in Section~\ref{sec:dist}. 
As the sample for the young population is too small, the constraints have a large uncertainty. The results are shown in Figure~\ref{fig:feh_young_vs_old_cov}. This plot was generated using the \texttt{pygtc} package \citep{Bocquet2016}.
We see that, for the younger population, an equal or lower metallicity is preferred in comparison to the one preferred for the older population.
However, the uncertainty is too high to yield a firm result.

\begin{figure*}[]
\centering
 {\includegraphics[scale=.8]{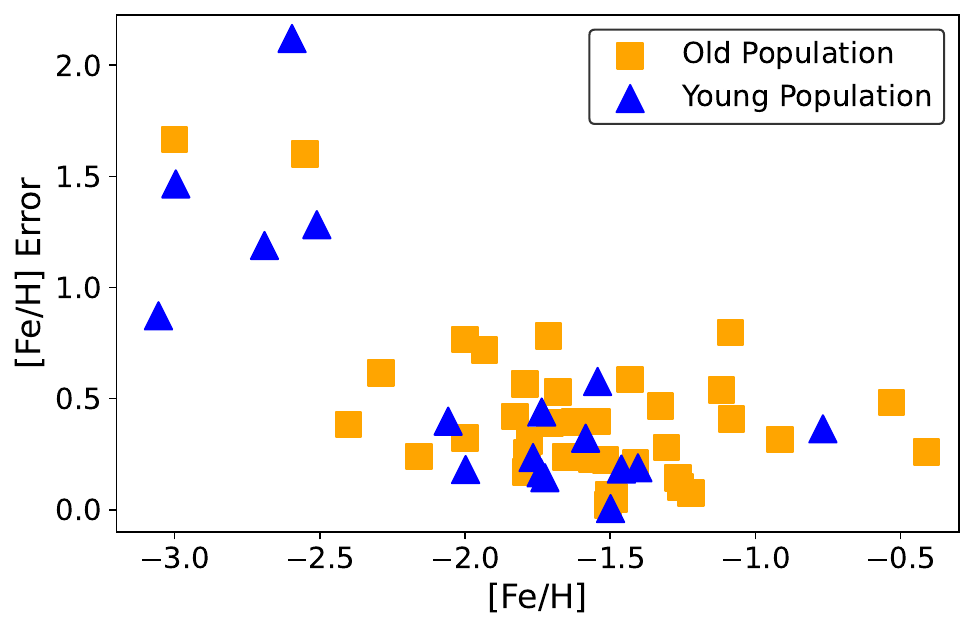}}
\caption{Plot of the measured [Fe/H] and [Fe/H] uncertainty for the sample of 55 stars, using \spexxy. The sample is divided into two, representing the young and old population in blue and orange, respectively.}
        \label{fig:feh_error}
\end{figure*}

In order to better understand the results, we plot the [Fe/H] measurements as a function of the uncertainty in Figure~\ref{fig:feh_error}.
We get some measurements that we consider to be outliers, with measured [Fe/H] $< -2.5$. These values, if real, are very surprising, in particular because most of them are for young stars. However, as the uncertainty is quite high for lower metallicity cases, it is not possible to have a definite conclusion. Furthermore, we repeated the MCMC analysis by removing the outliers and the results were unchanged, which means that, due to the large uncertainty, these measurements do not contribute significantly to the success of the fit. 
If these measurements are inaccurate, the outliers can be explained by the fact that these young stars barely have any metal lines in the MUSE spectral range. Therefore, \spexxy can wander off in the metallicity space, because changes in [Fe/H] do not affect the fit.
There is also the possibility that the parameter space is not well covered by the PHOENIX grid for these types of young stars, contributing to a poorer fit. 

\FloatBarrier

\FloatBarrier

\end{appendix}
\end{document}